\documentclass[fleqn,usenatbib]{mnras}

\usepackage{newtxtext,newtxmath}
\usepackage[T1]{fontenc}

\DeclareRobustCommand{\VAN}[3]{#2}
\let\VANthebibliography\thebibliography
\def\thebibliography{\DeclareRobustCommand{\VAN}[3]{##3}\VANthebibliography}

\usepackage{graphicx}	
\usepackage[flushleft]{threeparttable}
\usepackage{booktabs}
\usepackage{amsmath}
\usepackage{subfigure}
\usepackage{pdflscape}
\usepackage{array}
\usepackage{tabularx}
\usepackage{verbatim}
\usepackage{colortbl}
\usepackage{hyperref}
\usepackage{float}
\usepackage{adjustbox}
\usepackage{multirow}
\usepackage{xcolor}
\usepackage{soul}

\title[TOI-332\,b]{TOI-332\,b: a super dense Neptune found deep within the Neptunian desert}

\author[A. Osborn et al.]{
Ares~Osborn,$^{1,2}$,\thanks{E-mail: e.osborn@warwick.ac.uk}
David~J.~Armstrong$^{1,2}$,
Jorge~Fern\'andez~Fern\'andez$^{1,2}$,
Henrik~Knierim$^{3}$,
\newauthor
Vardan~Adibekyan$^{4}$,
Karen~A.~Collins$^{5}$,
Elisa~Delgado-Mena$^{4}$,
Malcolm~Fridlund$^{6}$,
Jo\~{a}o~Gomes~da~Silva$^{4}$,
\newauthor
Coel~Hellier$^{7}$,
David~G.~Jackson$^{8}$,
George~W.~King$^{9,1,2}$,
Jorge~Lillo-Box$^{10}$,
Rachel~A.~Matson$^{11}$,
\newauthor
Elisabeth~C.~Matthews$^{12}$,
Nuno~C.~Santos$^{4,13}$,
S\'ergio~G.~Sousa$^{4}$,
Keivan~G.~Stassun$^{14}$,
Thiam-Guan~Tan$^{15}$,
\newauthor
George~R.~Ricker$^{16}$,
Roland~Vanderspek$^{16}$,
David~W.~Latham$^{5}$,
Sara~Seager$^{16,17,18}$,
Joshua~N.~Winn$^{19}$,
\newauthor
Jon~M.~Jenkins$^{20}$,
Daniel~Bayliss$^{1,2}$,
Luke~G.~Bouma$^{21}$,
David~R.~Ciardi$^{22}$,
Kevin~I.~Collins$^{23}$,
\newauthor
Knicole~D.~Col\'{o}n$^{24}$,
Ian~J.~M.~Crossfield$^{16}$,
Olivier~D.~S.~Demangeon$^{4,13}$,
Rodrigo~F.~D\'iaz$^{25}$,
\newauthor
Caroline~Dorn$^{26}$,
Xavier~Dumusque$^{27}$,
Marcelo~Aron~Fetzner~Keniger$^{1,2}$,
Pedro~Figueira$^{27,4}$,
Tianjun~Gan$^{28}$,
\newauthor
Robert~F.~Goeke$^{16}$,
Andreas~Hadjigeorghiou$^{1,2}$,
Faith~Hawthorn$^{1,2}$,
Ravit~Helled$^{3}$,
Steve~B.~Howell$^{20}$,
\newauthor
Louise~D.~Nielsen$^{29}$,
Hugh~P.~Osborn$^{30}$,
Samuel~N.~Quinn$^{5}$,
Ramotholo~Sefako$^{31}$,
Avi~Shporer$^{16}$,
\newauthor
Paul~A.~Str{\o}m$^{1,2}$,
Joseph~D.~Twicken$^{32,20}$,
Andrew~Vanderburg$^{16}$,
and Peter~J.~Wheatley$^{1,2}$
\\
The authors' affiliations are shown in Appendix \ref{sec:affiliations}.
}

\date{Accepted XXX. Received YYY; in original form ZZZ}

\pubyear{2023}

\begin{document}
\label{firstpage}
\pagerange{\pageref{firstpage}--\pageref{lastpage}}
\maketitle

\begin{abstract}
To date, thousands of planets have been discovered, but there are regions of the orbital parameter space that are still bare. An example is the short period and intermediate mass/radius space known as the ``Neptunian desert’’, where planets should be easy to find but discoveries remain few. This suggests unusual formation and evolution processes are responsible for the planets residing here. We present the discovery of TOI-332\,b, a planet with an ultra-short period of $0.78$\,d that sits firmly within the desert. It orbits a K0 dwarf with an effective temperature of $5251 \pm 71$\,K. TOI-332\,b has a radius of $3.20^{+0.16}_{-0.12}$\,R$_{\oplus}$, smaller than that of Neptune, but an unusually large mass of $57.2 \pm 1.6$\,M$_{\oplus}$. It has one of the highest densities of any Neptune-sized planet discovered thus far at $9.6^{+1.1}_{-1.3}$\,g\,cm$^{-3}$. A 4-layer internal structure model indicates it likely has a negligible hydrogen-helium envelope, something only found for a small handful of planets this massive, and so TOI-332\,b presents an interesting challenge to planetary formation theories. We find that photoevaporation cannot account for the mass loss required to strip this planet of the Jupiter-like envelope it would have been expected to accrete. We need to look towards other scenarios, such as high-eccentricity migration, giant impacts, or gap opening in the protoplanetary disc, to try and explain this unusual discovery.
\end{abstract}




\section{Introduction}

\begin{figure*}
    \centering
    \includegraphics[width=\textwidth]{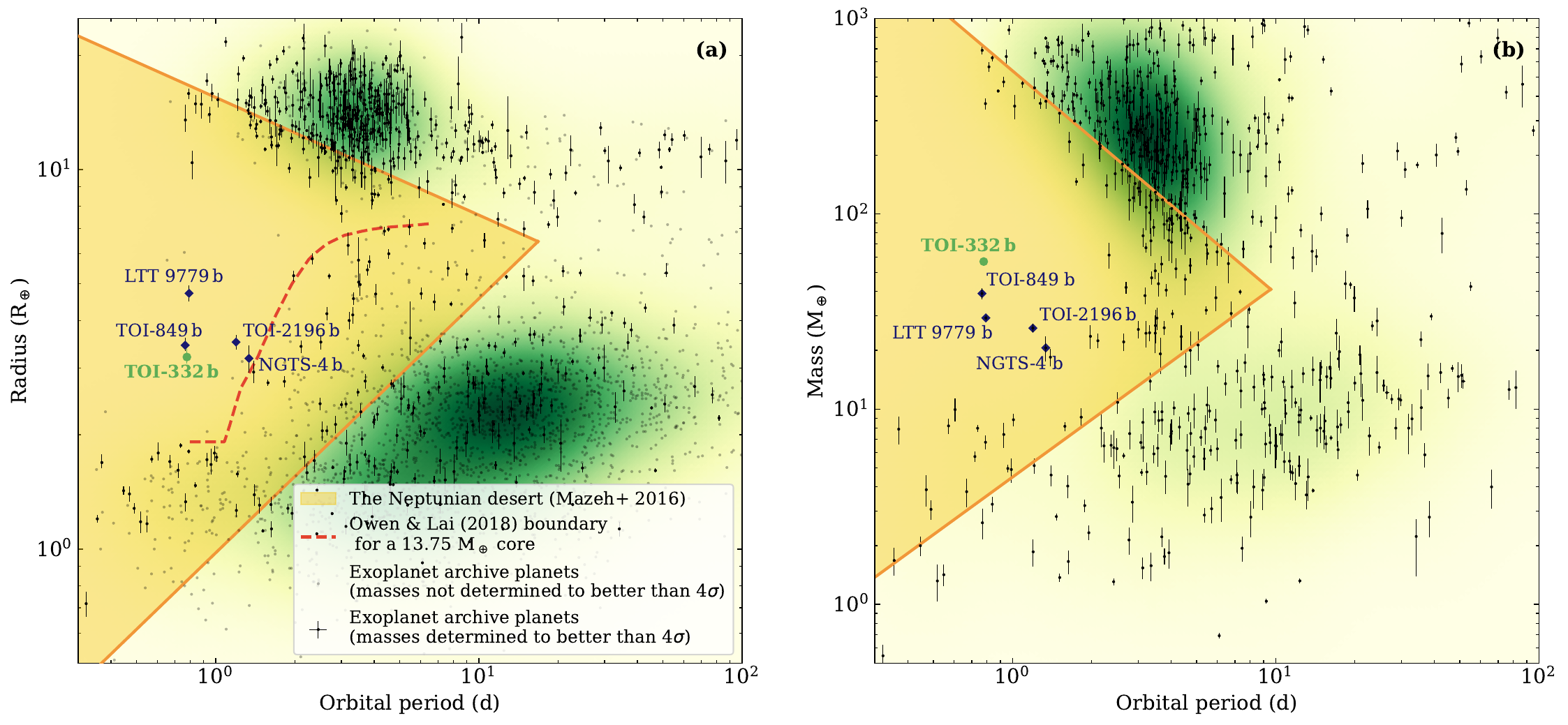}
    \caption{TOI-332\,b (green circle) in the context of the Neptunian desert, with \textbf{(a)} showing period-radius space, and \textbf{(b)} showing period-mass. The Neptunian desert boundaries from \citet{Mazeh2016} are plotted as solid lines, with the enclosed Neptunian desert area shaded in yellow. 
    In \textbf{(a)}, the dashed line is a numerical solution for the lower boundary of the desert determined for a 13.75\,M$_{\oplus}$ core in \citet{OwenLai2018}.
    Known planets were sourced from the NASA exoplanet archive (\url{https://exoplanetarchive.ipac.caltech.edu/}) on 3 May 2023: those without mass determinations or mass determinations worse than 4\,$\sigma$ are plotted as pale grey dots in \textbf{(a)} only; planets with mass determinations better than 4\,$\sigma$ are plotted as black dots in both \textbf{(a)} and \textbf{(b)}. 
    Population density of known planets is shaded in green, where darker green denotes more planets discovered in that region of parameter space: in \textbf{(a)} this includes all planets; in \textbf{(b)} this includes only planets with mass determination better than 4\,$\sigma$. 
    Particular planets with mass determination to better than 4\,$\sigma$ that are considered to be in the ``deep'' Neptunian desert are labelled (dark blue diamonds). 
    }
    \label{fig:nepdesert}
\end{figure*}

One of the key outcomes of the {\it Kepler} mission \citep{Borucki2010} was the population studies performed on the planets it discovered. This led to the identification of the ``Neptunian desert'' (also known as the ``hot Neptune desert'', ``sub-Jovian desert'', and ``evaporation desert''), a region of period-radius and period-mass parameter space where planets have, so far, been rarely found. The desert was first noted by \citet{SzaboKiss2011}, and has been the subject of many studies in the years since \citep[e.g.,][]{Boue2012,Beauge2013,Helled2016,Lundkvist2016}, and its boundaries were first formally defined by \citet{Mazeh2016}. As shown in Fig. \ref{fig:nepdesert}, it is a wedge shaped region where the upper boundary at large radii (or mass) decreases with increasing semi-major axis, and a lower boundary at small radii (or mass) which increases with increasing semi-major axis. The desert roughly encompasses intermediately-sized planets (approximately $2\,\textrm{R}_{\oplus} < R_p < 9\,\textrm{R}_{\oplus}$ and $10\,\textrm{M}_{\oplus} < M_p < 250\,\textrm{M}_{\oplus}$) with periods out to $\sim 5$\,days. 

This should not be due to an observational bias, as Neptune-sized planets with short periods are readily discovered by transit surveys like {\it Kepler} and, more recently, the Transiting Exoplanet Survey Satellite \citep[{\it TESS},][]{Ricker2015}. Theories have been put forward to explain the desert's existence and boundaries \citep[e.g.,][]{OwenLai2018,Vissapragada2022}. The lower boundary could be caused by photoevaporation of planets above the boundary, stripping their envelopes and reducing their radii/mass; while the upper boundary seems to be stable against photoevaporation, and may instead be understood as a ``tidal disruption barrier'', where planets below and left of the boundary migrating inwards can no longer successfully circularise and stabilise (see review by \citet{Dawson2018}).

In the years since its discovery, the desert has become more populated with planet discoveries, especially around its boundaries. However, there are so far only four planets with precisely determined masses (i.e. an error on their mass of better than 20 per cent) found deep within the desert, far from the boundaries set by \citet{Mazeh2016}: NGTS-4\,b \citep{West2019}; LTT-9779\,b \citep{Jenkins2020}; TOI-849\,b \citep{Armstrong2020}; and TOI-2196\,b \citep{Persson2022}. They are annotated in Fig. \ref{fig:nepdesert}. There are an additional few without precise masses: K2-100\,b \citep{Barragan2019}; K2-278\,b \citep{Livingston2018}, and Kepler-644\,b \citep{Morton2016}, the latter two being validated and having no mass determination. The few planets found in this barren desert are likely to have undergone unusual formation and/or evolutionary processes compared to those in more populated parameter spaces. There are now concerted efforts to find planets in and around the desert \citep[e.g.][]{Magliano2023,Bourrier2023} to determine what sculpts it. 

The aim of the HARPS-NOMADS program is to characterise planets in the Neptunian desert discovered by {\it TESS}, as the stars it observes are bright enough for effective radial velocity follow up. By precisely determining their masses and radii, we can constrain densities and thus the internal structures of these planets in order to understand their formation and evolution, leading to a better understanding of the origins of the desert itself. 

We present here the detection and characterisation of TOI-332\,b, an ultra-short period planet with an unusually high density located deep within the Neptunian desert. In Section\,\ref{observations}, we present the observations of the TOI-332 system, including photometry, spectroscopy, and high-resolution imaging. The spectroscopic analysis and derivation of chemical abundances of the star is then described in Section\,\ref{stellar}. In Section\,\ref{jointfit}, we describe the joint fit model to the data. In Section\,\ref{resultsdiscussion}, we present the results of the joint fit, discuss the nature of TOI-332\,b, theorise potential scenarios for its formation and evolution, and outline opportunities for further follow up of the system. Section\,\ref{conclusion} sets out our conclusions. 

\begin{table}
    \caption{Details for the TOI-332 system.}
	\label{tab:system}
	\begin{threeparttable}
	\begin{tabular}{llll}
	\toprule
	\textbf{Property}       & \textbf{(unit)}   & \textbf{Value}    & \textbf{Source} \\
	\midrule
	\multicolumn{4}{l}{\textbf{Identifiers}} \\
	TIC ID                  &                   & 139285832           & TICv8 \\
	2MASS ID                &                   & J23121409-4452349   & 2MASS \\
	Gaia ID                 &                   & 6529471108882243840 & Gaia DR3 \\
	\midrule
	\multicolumn{4}{l}{\textbf{Astrometric properties}} \\
	R.A.                    & (J2000.0)         & 23:12:14.10       & Gaia DR3 \\
	Dec                     & (J2000.0)         & -44:52:34.77      & Gaia DR3 \\
	Parallax                & (mas)             & 4.54 $\pm$ 0.03   & Gaia DR3 \\
	Distance                & (pc)              & 222.85 $\pm$ 3.69 & Gaia DR3 \\
	$\mu_{\rm{R.A.}}$       & (mas\,yr$^{-1}$)  & 35.86 $\pm$ 0.01  & Gaia DR3 \\
	$\mu_{\rm{Dec}}$        & (mas yr$^{-1}$)   & -37.62 $\pm$ 0.02 & Gaia DR3 \\
	\midrule
	\multicolumn{4}{l}{\textbf{Photometric properties}} \\
	{\it TESS}              & (mag)             & 11.527 $\pm$ 0.006 & TICv8 \\
	B                       & (mag)             & 13.10 $\pm$ 0.03   & TICv8 \\
	V                       & (mag)             & 12.35 $\pm$ 0.05   & TICv8 \\
	G                       & (mag)             & 12.0545 $\pm$ 0.0002 & Gaia DR3 \\
	J                       & (mag)             & 10.78 $\pm$ 0.02   & 2MASS \\
	H                       & (mag)             & 10.41 $\pm$ 0.02   & 2MASS \\
	K                       & (mag)             & 10.32 $\pm$ 0.02   & 2MASS \\
	\bottomrule
	\end{tabular}
	\begin{tablenotes}
	\item \textbf{Sources:} TICv8 \citep{Stassun2019}, 2MASS \citep{Skrutskie2006}, Gaia Data Release 3 \citep{GaiaDR3}.
	\end{tablenotes}
	\end{threeparttable}
\end{table}

\section{Observations}\label{observations}

\begin{figure}
    \centering
    \includegraphics[width=\columnwidth]{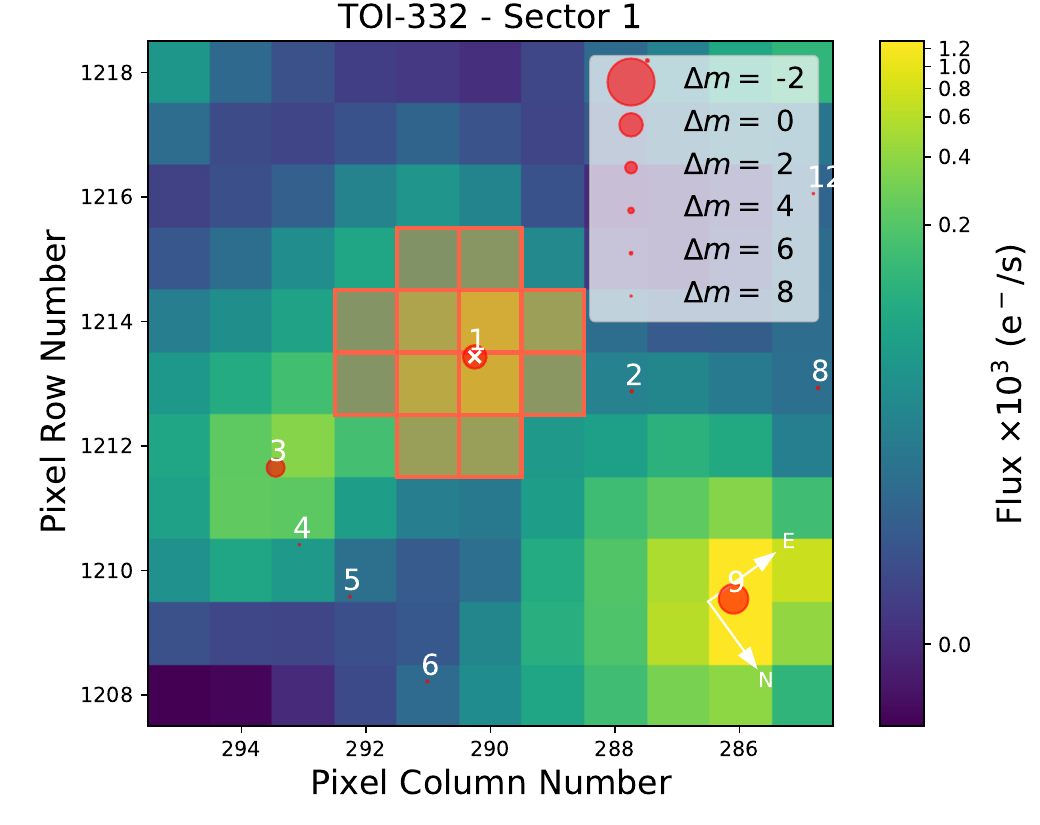}
    \caption{The Target Pixel File (TPF) for TOI-332 (marked as a white cross) from {\it TESS} S1. Other {\it Gaia} DR3 sources within a limit of 8 {\it Gaia} magnitudes difference from TOI-332 are marked as red circles, and are numbered in distance order from TOI-332. The aperture mask is outlined and shaded in red. This figure was created with {\tt tpfplotter} \citep{Aller2020}.}
    \label{fig:tpf}
\end{figure}

In this section, we describe the instrumentation and observations used for the detection and characterisation of the TOI-332 system.

\begin{figure*}
    \centering
    \includegraphics[width=\textwidth]{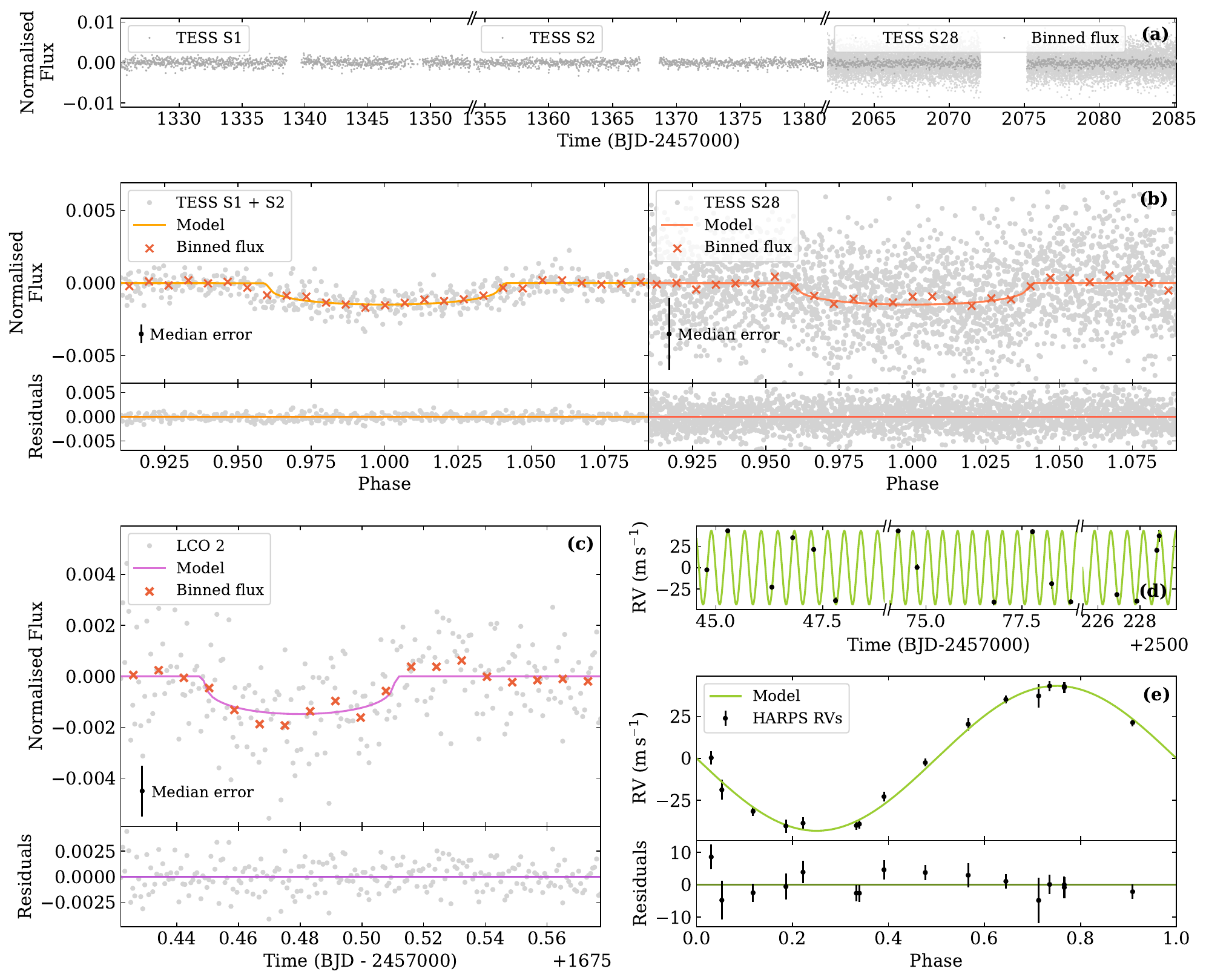}
    \caption{Joint fit model to the {\it TESS}, LCO, and HARPS data. \\
    \textbf{(a)} {\it TESS} PDCSAP light curve for Sectors 1, 2 and 28 (circles), with time given as Barycentric Julian Date (BJD). Sectors 1 and 2 are 30\,min cadence data, Sector 28 is 2\,min cadence, hence the higher levels of noise in the latter. The Sector 28 data binned to 30\,min has been overplotted in dark grey.\\
    \textbf{(b)} Phase folded {\it TESS} 30\,min cadence data (grey circles) from Sectors 1 and 2 (left), and 2\, min cadence data from Sector 28 (right). Binned flux (red crosses) and the best fit model (solid line) are overplotted, and the median error on the flux is displayed (one standard deviation, black error bar, bottom left). Residuals when the model is subtracted are shown in the bottom panels. \\
    \textbf{(c)} An example of the phase folded LCO data, using  the second transit obtained (chronologically) by LCO (the model is fit to all of the LCO transits and the full data can be seen in Fig.\,\ref{fig:lco}). Symbols and model are as in \textbf{(b)}, with residuals in the bottom panel. \\
    \textbf{(d-e)} The HARPS data (black circles), shown as a time series in \textbf{(d)}, and the phase folded data in \textbf{(e)}. The model is plotted as in \textbf{(b)}, with residuals in the bottom panel.
    } 
    \label{fig:jointfit}
\end{figure*}

\subsection{Photometry}

\subsubsection{\it TESS}\label{tess}

The TOI-332 system (TIC\,139285832, see Table \ref{tab:system}) was observed in {\it TESS} Sectors 1 (25 July - 22 Aug 2018, hereafter S1) and 2 (22 August - 20 September 2018, hereafter S2) with a 30\,min cadence in the full-frame images (FFIs). TOI-332.01 (now TOI-332\,b) was detected in the FFIs by the MIT Quick-Look Pipeline \citep[QLP,][]{Huang2020} as part of the early {\it TESS} Data Alerts, and alerted on 20 December 2018 \citep{Guerrero2021}. It was then re-observed in Sector 28 (30 July  - 26 August 2020, hereafter S28) on Camera 2 with a 2\,min cadence. The Data Validation report \citep{Twicken2018,Li2019} difference image centroid offsets determined from the S28 pixel data locate the transit source within $2.56 \pm 2.76$\,arcsec of TOI-332, and exclude all other TICv8 objects as possible sources of the transit signal. The detection gave a period of $0.77685 \pm 0.0003$\,d, a transit duration of $1.43 \pm 0.442$\,h, and a depth of $830 \pm 8$\,ppm. The data products, including calibrated full-frame images and light curves, are available on the Mikulski Archive for Space Telescopes (MAST; \url{https://archive.stsci.edu/missions-and-data/transiting-exoplanet-survey-satellite-tess}), and were produced by the {\it TESS} Science Processing Operations Center \citep[SPOC,][]{Jenkins2016,Caldwell2020} at NASA Ames Research Center. 

We downloaded the publicly available photometry provided by the SPOC pipeline, and used the Presearch Data Conditioning Simple Aperture Photometry (PDCSAP), from which common trends and artefacts have been removed by the SPOC Presearch Data Conditioning (PDC) algorithm \citep{Twicken2010,Smith2012,Stumpe2012,Stumpe2014}. The median-normalised PDCSAP flux, after removal of data points flagged as being affected by excess noise, is shown in Fig.\,\ref{fig:jointfit}. No further detrending of the light curves was deemed necessary as they are relatively flat across the whole time series, showing little stellar activity. We also recover no periodicity from either the PDCSAP or SAP (Simple Aperture Photometry, where no trends and artifacts have been removed) flux that may be indicative of a stellar rotation period. The phase folded transits and best fit model are also shown in Fig.\,\ref{fig:jointfit}.

\subsubsection{LCOGT}\label{lcogt} 

The {\it TESS} pixel scale is $\sim$\,21\,arcsec per pixel and photometric apertures typically extend out to roughly 1 arcmin, generally causing multiple stars to blend in the {\it TESS} aperture (the aperture for the {\it TESS} S1 data for TOI-332 is shown in Fig.\,\ref{fig:tpf}). To attempt to determine the true source of the TOI-332 detection in the {\it TESS} data and refine its ephemeris and transit shape, we conducted ground-based photometric follow-up observations of the field around TOI-332 as part of the {\it TESS} Follow-up Observing Program\footnote{https://tess.mit.edu/followup} Sub Group 1 \citep[TFOP;][]{collins2019}.

We observed six full predicted transit windows of TOI-332.01 using the Las Cumbres Observatory Global Telescope \citep[LCOGT;][]{Brown2013} 1.0\,m network nodes. The details of each observation are provided in the caption of Fig.\,\ref{fig:lco}. We used the {\tt TESS Transit Finder}, which is a customized version of the {\tt Tapir} software package \citep{Jensen2013}, to schedule our transit observations. The 1\,m telescopes are equipped with $4096\times4096$ SINISTRO cameras having an image scale of 0.389\,arcsec per pixel, resulting in a 26\,arcsec\,$\times$\,26\,arcsec field of view. The images were calibrated by the standard LCOGT {\tt BANZAI} pipeline \citep{McCully2018}. Differential photometric data were extracted using {\tt AstroImageJ} \citep{Collins2017}. As shown in Fig.\,\ref{fig:lco}, we detected transit-like signals in all six TOI-332 follow-up light curves using photometric apertures with radii in the range of 3.1\,arcsec to 7.8\,arcsec, which exclude flux from the nearest neighbour of TOI-332 in the {\it Gaia} DR3 and TICv8 catalogs (which is 51\,arcsec northeast, and is the target numbered 2 in Fig.\,\ref{fig:tpf}). We therefore confirm that the TOI-332.01 signal in the {\it TESS} data occurs on-target relative to all known {\it Gaia} DR3 and TICv8 stars.

\subsubsection{PEST}\label{pest} 

We observed TOI-332 in the V band from the Perth Exoplanet Survey Telescope (PEST) near Perth, Australia. At the time, the 0.3\,m telescope was equipped with a $1530\times1020$ SBIG ST-8XME camera with an image scale of 1.2\,arcsec per pixel resulting in a 31\,arcsec\,$\times$\,21\,arcsec field of view. A custom pipeline based on C-Munipack was used to calibrate the images and extract the differential photometry. Unfortunately, there is a gap during the transit egress due to cloud cover, and poor weather negatively affected the data quality; therefore, we do not include it in our joint fit model, but present the data with the model over-plotted in Fig.\,\ref{fig:pest}. 

\subsubsection{WASP}\label{wasp} 

WASP-South, an array of 8 wide-field cameras, was the Southern station of the WASP transit-search project \citep{Pollacco2006}. It observed the field of TOI-332 in the years 2006, 2007, 2010 and 2011 when equipped with 200-mm, f/1.8 lenses, and then again in 2012, 2013 and 2014 when equipped with 85-mm, f/1.2 lenses. It observed on each clear night, with a typical 10\,min cadence, and accumulated 88\,000 photometric data points on TOI-332.

We searched the data for any rotational modulation using the methods from \citet{Maxted2011}. We find a significant periodicity in data from one year, spanning 168 nights in 2007, with an estimated false-alarm probability of 0.15\,per\,cent. The period is $20.9 \pm 1.0$\,d and the amplitude 3\,mmag. We note that there is also a peak near 36 days, though it is not significant in itself. In 2012, a possible modulation with a similar period ($18.4 \pm 1.5$\,d) has a lower significance (8\,per\,cent false-alarm likelihood). No significant periodicity was detected in other years. We discuss the periodicity further in Section\,\ref{stellar}. 


We also note that the standard WASP transit-detection algorithm \citep{CollierCameron2007}, when run on the same 2007 dataset, detects the transit and reports a period of $0.77663 \pm 0.00012$\,d with an epoch of TDB $2454343.4652 \pm 0.0079$. The period matches the {\it TESS} period to 1 part in 2000 while the phase matches an extrapolated ephemeris to within 3\,per\,cent. However, when run on the full dataset combined, the algorithm does not find the transit, though this is explainable given that red noise in other years can destroy the detection. At a depth of 0.15\,per\,cent, the transit would be the shallowest detected in WASP data, though it is comparable to the detection of the 0.17\,per\,cent transit of HD\,219666\,b \citep{Hellier2019}. We conclude that this detection is likely, but not securely, real, and thus we report it here as the earliest detection of the transit of TOI-332\,b. Since planets in short-period orbits are expected to undergo tidal decay, timings over the longest possible time span are of interest, and we discuss this further in Section\,\ref{orbdecay}. 

Due to the uncertainty in the detection, we do not include the WASP data in our joint fit model, but we present the 2007 dataset with the best fit model over-plotted in Fig.\,\ref{fig:wasp}. 

\begin{figure}
    \centering
    \includegraphics[width=\columnwidth]{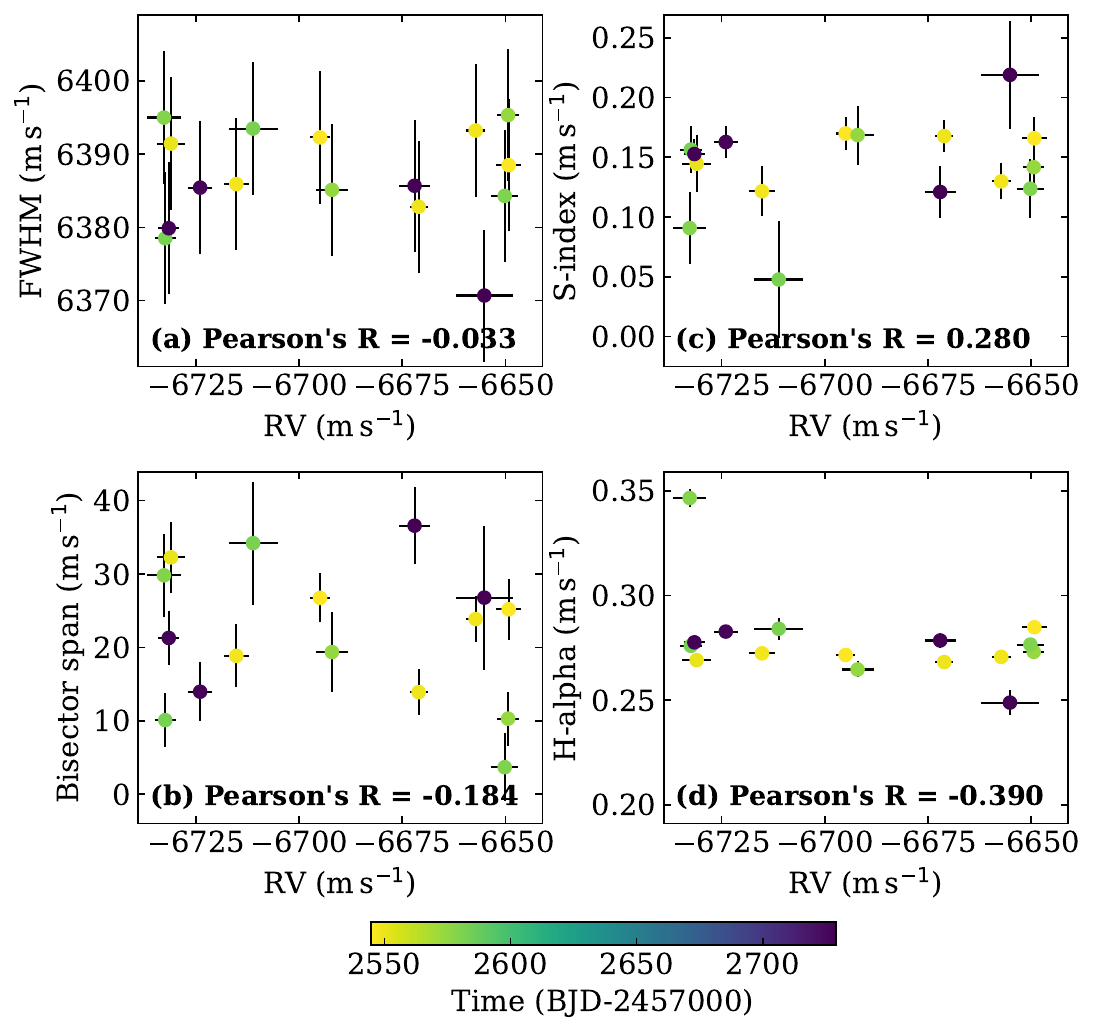}
    \caption{HARPS radial velocities plotted against stellar activity indicators: \\ \textbf{(a)}, the full-width at half-maximum (FWHM) of the cross-correlation function (CCF); \\
    \textbf{(b)}, the bisector span of the CCF; \\
    \textbf{(c)}, the s-index; \\
    and \textbf{(d)}, h-alpha. \\
    The Pearson's R statistic, a measure of correlation strength, is given for each, and no significant correlation is seen. Colour represents the time of observation in Barycentric Julian Date (BJD). All error bars show one standard deviation.}
    \label{fig:rvcorr}
\end{figure}

\subsection{Spectroscopy}

\subsubsection{HARPS}\label{harps}

We made radial velocity (RV) measurements of TOI-332  with the High Accuracy Radial velocity Planet Searcher (HARPS) spectrograph mounted on the ESO 3.6\,m telescope at the La Silla Observatory in Chile \citep{Pepe2002}. A total of 16 spectra were obtained between 25 November 2021 and 29 May 2022 under the HARPS-NOMADS large programme (ID 1108.C-0697, PI: Armstrong). The instrument (with resolving power $R = 115\,000$) was used in high-accuracy mode (HAM) with an exposure time of 2400\,s, and 1-2 observations of the star were made per night. The data were reduced using the standard offline HARPS data reduction pipeline, and a K5 template was used in a weighted cross-correlation function (CCF) to determine the RV values \citep{Baranne1996,Pepe2002}. The line bisector (BIS) and full-width at half-maximum (FWHM) were measured using previously published methods \citep{Boisse2011}. The RV measurements can be found in Table \ref{tab:rvs}, and the RV data and single-planet Keplerian best fit are shown in Fig.\,\ref{fig:jointfit}. 

\begin{figure}
    \centering
    \includegraphics[width=\columnwidth]{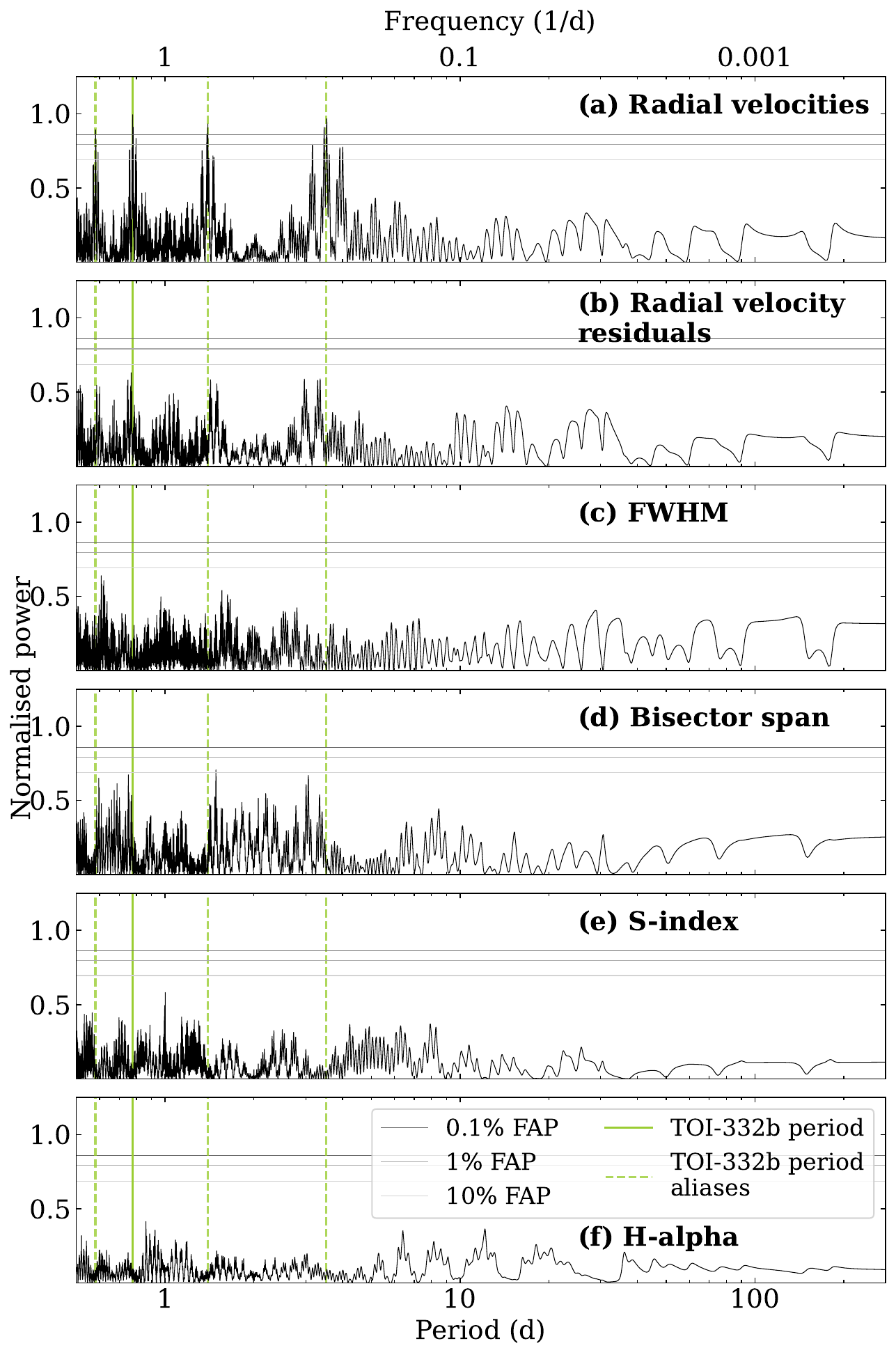}
    \caption{Periodograms for the HARPS data. The expected period of TOI-332\,b is denoted by a solid vertical line, with the aliases of this period given as dashed lines. The 0.1, 1, and 10 per cent False Alarm Probabilities (FAPs) are shown as solid horizontal lines. The FAPs are calculated using the approximation from \citet{Baluev2008}. 
    From top to bottom: \\
    \textbf{(a)} the periodogram for the raw radial velocities with a peak above the 0.1 per cent FAP at the expected planetary period; \\
    \textbf{(b)} the periodogram for the radial velocity residuals after the best fit model has been removed, showing no further significant peaks; \\
    \textbf{(c)-(f)} the periodograms for the stellar activity indicators full-width at half-maximum (FWHM, \textbf{(c)}), bisector span \textbf{(d)}, s-index \textbf{(e)}, and h-alpha \textbf{(f)}, with no significant periodicity shown.}
    \label{fig:rvperiodogram}
\end{figure}

No correlation was detected between the RVs and the FWHM and bisector span of the CCF, or the S and H$\alpha$ activity indexes, shown in Fig.\,\ref{fig:rvcorr}. After removing the contribution from TOI-332\,b, we studied the RV residuals and found no evidence of further periodicity in those or in the activity indicators as shown in Fig.\,\ref{fig:rvperiodogram}. Unfortunately, this means the RVs give no indication of a possible stellar rotation period that would corroborate that found by WASP in Section\,\ref{wasp} or derived later in Section\,\ref{stellar}. We note that one of the H$\alpha$ points, from the night of 29 December 2021, is an outlier, shown in Fig.\,\ref{fig:rvcorr}. Investigating the spectrum from this night, we find a narrow emission line in the centre of the H$\alpha$ line that may be indicative of a flare; however, the RV point corresponding to this night is not an outlier, nor does it have an anomalously large error, so we retain it.

\subsection{High resolution imaging}

\begin{figure}
    \centering
    \begin{subfigure}
        \centering
        \includegraphics[width=\columnwidth]{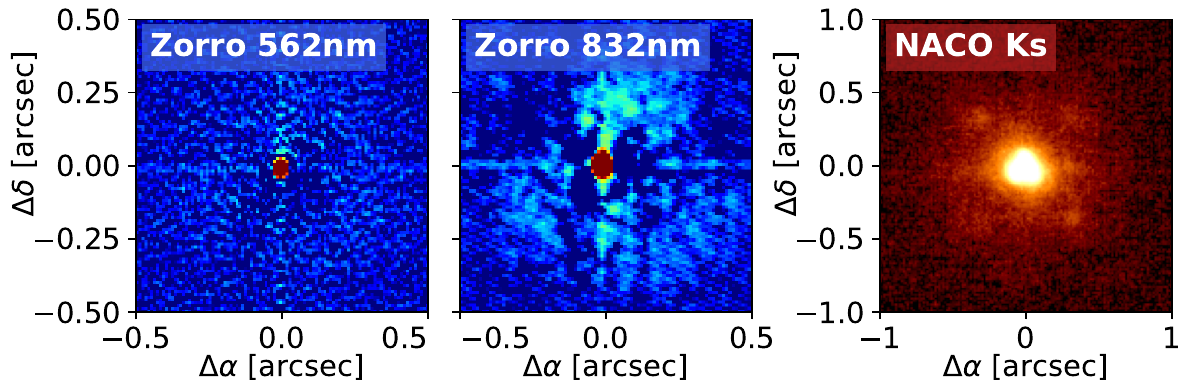}
    \end{subfigure}
    \begin{subfigure}
        \centering
        \includegraphics[width=\columnwidth]{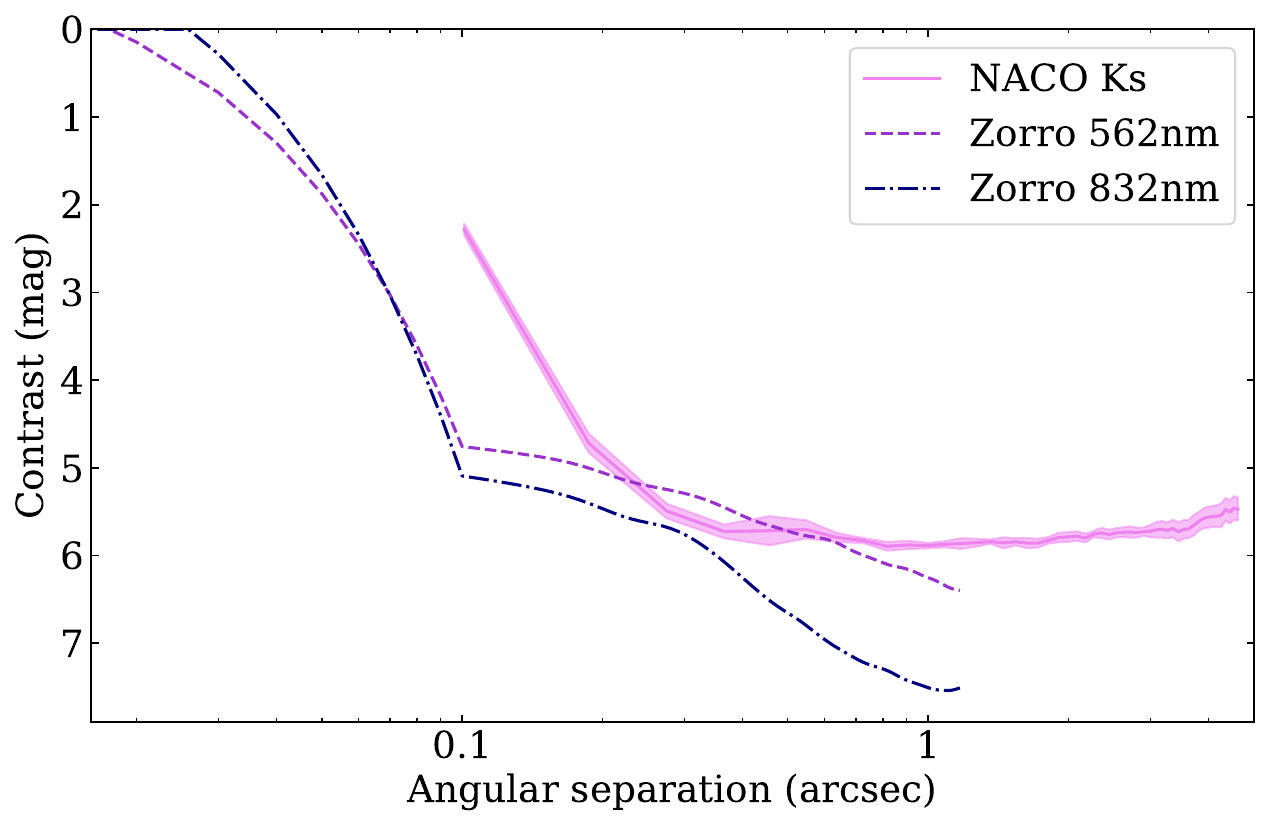}
    \end{subfigure}
    \caption{A compilation of reconstructed images for the sources of high-resolution imaging described in Sections\,\ref{gemini} and \ref{vlt} (top), with their corresponding 5$\sigma$ contrast curves (bottom). Zorro observes simultaneously in the 562 and 832\,nm bands; NACO observes in the near-IR with a Ks filter (labelled). No additional companions are detected.}
    \label{fig:hrimaging}
\end{figure}

\subsubsection{Gemini Zorro}\label{gemini} 

High-angular resolution images of TOI-332 were obtained on 10 October 2019 using the Zorro\footnote{https://www.gemini.edu/sciops/instruments/alopeke-zorro/} speckle instrument on the Gemini-South telescope \citep{Scott2021}. Zorro observes simultaneously in two bands ($832 \pm 40$\,nm and $562 \pm 54$\,nm), obtaining diffraction limited images with inner working angles of 0.026 and 0.017\,arcsec respectively. The TOI-332 data set consisted of 5 sets of $1000 \times 0.06$\,s images, which were combined using Fourier analysis techniques, examined for stellar companions, and used to produce reconstructed speckle images \citep[see][]{Howell2011}. The speckle imaging reveals TOI-332 to be a single star with no companions detected within 1.2\,arcsec down to contrast limits of $\sim5-7$ mag, shown in Fig.\,\ref{fig:hrimaging}. At the distance of TOI-332 (220\,pc), these angular limits correspond to spatial limits of $4 - 264$\,au. 

\subsubsection{VLT NaCo}\label{vlt} 

We collected high-resolution adaptive optics imaging of TOI-332 with VLT/NaCo on 19 June 2019. These near-IR images complement the visible-band speckle data and provide greater sensitivity to late-type bound companions. We collected a sequence of nine images in the Ks filter, each with an integration time of 11\,s; the telescope was dithered between each exposure and a sky background frame was created by median combining the science frames. We removed bad pixels, flat fielded, subtracted the sky background, and aligned the images on the stellar position before co-adding the sequence. We calculated the sensitivity of these images by injecting fake companions at several position angles and separations, and measuring the significance to which they could be recovered.

The AO image and the sensitivity limits are presented in Fig.\,\ref{fig:hrimaging}. Some extended PSF structure is seen in the image, but the star is unresolved and no companions are identified. The data reveal that TOI-332 is a single star down to 5.5\,mag of contrast, beyond 365\,mas from the star, and exclude bright companions beyond 100\,mas.

\section{Spectroscopic analysis and chemical abundances}\label{stellar}


Here we perform several different methods to measure and derive a range of stellar parameters for TOI-332. 

We first used ARES+MOOG to derive spectroscopic stellar parameters ($T_{\mathrm{eff}}$, $\log g$, microturbulence $v_{\rm tur}$, and [Fe/H]) following the same methodology as described in \citet{Santos2013,Sousa2014,Sousa2021}. The latest version of ARES \footnote{The latest version, ARES v2, can be downloaded at https://github.com/sousasag/ARES} \citep{Sousa2007,Sousa2015} was used to consistently measure the equivalent widths (EW) of selected iron lines in the combined spectrum of TOI-332. For this, we used the iron line list presented in \citet{Sousa2008}. The best spectroscopic parameters are found by converging into ionisation and excitation equilibrium. This process makes use of a grid of Kurucz model atmospheres \citep{Kurucz1993} and the radiative transfer code MOOG \citep{Sneden1973}. We also derived a  trigonometric surface gravity using {\it Gaia} DR3 data following the same procedure as described in \citet{Sousa2021}. We find values of: $T_{\mathrm{eff}} = 5251 \pm 71$\,K; $\log g = 4.46 \pm 0.04$\,c\,g\,s; $v_{\rm tur} = 0.815 \pm 0.069$\,km\,s$^{-1}$; and ${\rm [Fe/H]} = 0.256 \pm 0.048$\,dex.

To estimate the stellar mass and radius we used the calibrations in \citet{Torres2010}; because the mass is between 0.7 and 1.3\,$M_\odot$ we used the correction in \citet{Santos2013}. This gives $R_\star = 0.87 \pm 0.03$\,R$_{\odot}$ and $M_\star = 0.88 \pm 0.02$\,M$_{\odot}$. 

Stellar abundances of the elements were then derived using the classical curve-of-growth analysis method assuming local thermodynamic equilibrium. The same codes and models were used for the abundance determinations. For the derivation of chemical abundances of refractory elements we closely followed the methods described in \citep[e.g.][]{Adibekyan2012, Adibekyan2015, DelgadoMena2017}. Abundances of the volatile elements, C and O, were derived following the method of \cite{DelgadoMena2021, Bertrandelis2015}. Since the two spectral lines of oxygen are usually weak and the 6300.3\AA{} line can be contaminated by tellurics or an oxygen airglow, the EWs of these lines were manually measured with the task \texttt{splot} in IRAF. All the [X/H] ratios are obtained by doing a differential analysis with respect to a high S/N solar (Vesta) spectrum from HARPS. The abundances of these elements are presented in Table \ref{tab:abundances}. 

Under the assumption that stellar composition serves as a reliable indicator of the disc composition during the planet formation phase, we can determine the mass fraction of the planet building blocks. Following the methodology outlined in \citet{Santos2015,Santos2017}, which uses a simple stoichiometric model and chemical abundances of Fe, Mg, and Si, we computed that the anticipated iron-to-silicates mass fraction is $33.5 \pm 3.1$\,per\,cent.


To estimate the activity level of TOI-332, we used the HARPS spectra to calculate the $\log R'_\mathrm{HK}$ activity index. We co-added all spectra and used ACTIN2\footnote{Available at \url{https://github.com/gomesdasilva/ACTIN2}.} \citep{GomesdaSilva2018, GomesdaSilva2021} to extract the $S_\mathrm{CaII}$ index. This index was calibrated to the Mt. Wilson scale using the calibration in \citet{GomesdaSilva2021} and converted to $\log R'_\mathrm{HK}$ via \citet{Noyes1984}, giving $\log R'_\mathrm{HK} = -4.831 \pm 0.003$. This can then be used to derive a rotation period ($P_{\rm rot}$) and age of the star ($\tau$) via the relations in \citet{Mamajek2008}, giving a rotation period of $35.6 \pm 4.6$\,d and an age of $5.0 \pm 2.3$\,Gyr. This rotation period is approximately twice those obtained by WASP in Section\,\ref{wasp}, and so the WASP detection could be the first harmonic rather than the true rotational period. 

Moreover, we used the chemical abundances of some elements to derive an alternative value for the age through the so-called chemical clocks (i.e. certain chemical abundance ratios which have a strong correlation for age). We applied the 3D formulas described in Table 10 of \citet{DelgadoMena2019}, which also consider the variation in age produced by the effective temperature and iron abundance. The chemical clocks [Y/Mg], [Y/Zn], [Y/Ti], [Y/Si], [Y/Al], [Sr/Ti], [Sr/Mg] and [Sr/Si] were used from which we obtain a weighted average age of $6.3 \pm 1.8$\,Gyr. This age is in agreement (within errors) with the age obtained from the stellar activity and rotation.

As an independent determination of the basic stellar parameters, we performed an analysis of the broadband spectral energy distribution (SED) of the star together with the {\it Gaia} EDR3 parallax \citep[with no systematic offset applied; see, e.g.,][]{StassunTorres2021}, in order to determine an empirical measurement of the stellar radius, following the procedures described in \citet{StassunTorres2016,Stassun2017,Stassun2018}. We pulled the $JHK_S$ magnitudes from {\it 2MASS}, the W1--W3 magnitudes from {\it WISE}, the $G_{\rm BP} G_{\rm RP}$ magnitudes from {\it Gaia}, and the NUV magnitude from {\it GALEX}. Together, the available photometry spans the full stellar SED over the wavelength range 0.2--10~$\mu$m (see Fig.\,\ref{fig:sed}).

We performed a fit using Kurucz stellar atmosphere models, with the effective temperature ($T_{\mathrm{eff}}$), surface gravity ($\log g$), and metallicity ([Fe/H]) adopted from the spectroscopic analysis above. The remaining free parameter is the extinction $A_V$, which we limited to the maximum line-of-sight value from the Galactic dust maps of \citet{Schlegel1998}. The resulting fit (Fig.\,\ref{fig:sed}) has a reduced $\chi^2$ of 1.3, excluding the {\it GALEX} NUV flux which indicates a moderate level of activity (see below), and a best fit $A_V = 0.02 \pm 0.02$. Integrating the (unreddened) model SED gives the bolometric flux at Earth, $F_{\rm bol} = 3.851 \pm 0.045 \times 10^{-10}$ erg~s$^{-1}$~cm$^{-2}$. Taking the $F_{\rm bol}$ and $T_{\mathrm{eff}}$ together with the {\it Gaia} parallax gives the stellar radius, $R_\star = 0.923 \pm 0.016$~R$_\odot$. In addition, we can estimate the stellar mass from the empirical relations of \citet{Torres2010}, giving $M_\star = 0.96 \pm 0.06$~M$_\odot$. These broadly agree with the previous values.

Finally, to obtain another independent check on the fundamental stellar parameters, and following \citet{Fridlund2020} and references therein, we analysed our spectrum with version 5.22 of the spectral analysis package \href{http://www.stsci.edu/~valenti/sme.html}{{\tt{SME}}} \citep[Spectroscopy Made Easy;][]{ValentiPiskunov1996,PiskunovValenti2017}. This IDL based software is used to fit the observations to synthetic stellar spectra calculated with a given set of input parameters and a suitable atmospheric grid. Here, we used the Atlas12 \citep{Kurucz2013} grids, together with atomic and molecular line data from \href{http://vald.astro.uu.se}{VALD} \citep{Ryabchikova2015} to calculate the synthetic spectra. For $T_\mathrm{eff}$, we modelled the line wings of the hydrogen alpha line, and derived the surface gravity, $\log g$, from the calcium triplet $\lambda$6102, 6122, and 6162, and the $\lambda$6439 line. For an independent check, we also modelled the Na I doublet at 5888/89\,\AA. We find $T_\mathrm{eff} = 5185 \pm 100$\,K and $\log g = 4.4 \pm 0.1$, both in agreement with the values determined using ARES+MOOG. 

We then fitted a large number of iron lines to obtain the abundances ${\rm [Fe/H]} = 0.4 \pm 0.1$\,dex; ${\rm [Ca/H]} = 0.47 \pm 0.1$\,dex; and ${\rm [Na/H]} = 0.47 \pm 0.1$\,dex. 

Following again schemes described in \citet{Fridlund2020} and keeping the macroturbulent $v_\mathrm{mac}$~and microturbulent $v_\mathrm{mic}$~velocities fixed at the empirical values found in the literature \citep{Bruntt2010, Doyle2014}, we find $v \sin i_\star = 1.5 \pm 1.2$\,km\,s$^{-1}$. We can use the rotational period of 35.6\,d derived earlier to estimate an equatorial velocity of $\approx 1.24$\,km\,s$^{-1}$ (assuming spin-orbit alignment) which is in agreement with this, supporting the hypothesis of the WASP period being half the true period. We therefore take forward the 35.6\,d stellar rotation period into our later analysis. 

There are uncertainties on the values for $R_\star$ and $M_\star$ due to the methods used to derive them: for example, in the calculation of the synthetic models used to fit the observed spectra, and in the \citet{Torres2010} calibration used. Errors on the primary derived stellar parameters ($T_{\mathrm{eff}}$, $\log g$, ${\rm [Fe/H]}$) are taken into account when applying the \citet{Torres2010} calibration, as explained in \citet{Santos2013}.

We have used multiple methods to derive stellar parameters to account for unknown systematic effects. Our results from each method are consistent, implying that our stated errors are reasonable and the effect of unknown systematics is small. We note that any systematic errors remaining will propagate into the planetary parameters.

\section{The joint fit}\label{jointfit}

Using the {\tt exoplanet} package \citep{exoplanet:exoplanet}, we fit the photometry from {\it TESS} and LCOGT simultaneously with the RVs from HARPS. {\tt exoplanet} utilises the light curve modelling package {\tt Starry} \citep{exoplanet:luger18}, {\tt PyMC3} \citep{exoplanet:pymc3}, and {\tt celerite} \citep{exoplanet:celerite}. For consistency, all timestamps were converted to the same time system, that used by {\it TESS}, i.e. BJD\,-\,2457000 (BJD-TDB). All prior distributions set on the parameters fit in this model are given in Table \ref{tab:jointfit}. 

The photometric flux is normalised by dividing the full individual light curves by the median of their out-of-transit points and subtracting unity to produce a lightcurve with out-of-transit flux of zero. No further detrending is deemed to be necessary for either the LCOGT or {\it TESS} data, and so none is included in the joint fit. 

To model the planetary transits, we use a limb-darkened transit model utilising the quadratic limb-darkening parameterisation in \citet{exoplanet:kipping13} and a Keplerian orbit model. We put Gaussian priors informed by the ARES+MOOG values on the stellar radius $R_\star$ and the stellar mass $M_\star$.

The Keplerian orbit model is parameterised for the planet in terms of the orbital period $P$, the time of a reference midtransit $t_c$, the eccentricity $e$, and the argument of periastron $\omega$. In an earlier iteration of this model, we found the eccentricity of TOI-332\,b to be consistent with 0 (with the 95 per cent confidence interval for the eccentricity being 0 to 0.15), and so fix $e$ and $\omega$ to 0 in the final model presented here. A close-to-zero eccentricity is also expected given the very short orbital period. These parameters are then input into light curve models created with {\tt Starry}, alongside further parameters which are planetary radii $R_p$, the time series of the data $t$, and the exposure time $t_{\mathrm{exp}}$ of the instrument. 

Individual light curve models are created for the LCOGT data, the combined {\it TESS} S1 and S2 data, and the {\it TESS} S28 data (S28 is kept separate to S1 and S2 due to differing cadence and exposure time of the S1 and S2 data compared to the S28 data). We use values from the {\it TESS} SPOC pipeline \citep{Li2019} to estimate the placement of wide, uninformative uniform priors for the epoch, period, and radius of TOI-332\,b. For each lightcurve, we put a Gaussian prior on the offset with a mean of zero and standard deviation of one.

\begin{table*}
    \small
    \caption{Stellar parameters of TOI-332, and transit, orbital, and physical parameters of TOI-332\,b (further parameters from the joint fit model can be found in Table \ref{tab:jointfit}).}
	\label{tab:params}
	\begin{threeparttable}
	\begin{tabular}{llll}
	\toprule
	\textbf{Parameter}                      & \textbf{(unit)}           & \textbf{Value}        & \textbf{Source} \\
	\midrule
	\multicolumn{4}{l}{\textbf{Host star}} \\
	Distance to Earth                       & (pc)                      & $222.85 \pm 3.69$     & Gaia DR3 \\
	Effective temperature $T_{\rm eff}$     & (K)                       & $5251   \pm 71  $     & ARES+MOOG \\
    Spectral type                           & -                         & K0V                   & \citet{Pecaut2013} \\
	Surface gravity $\log g$                & (c\,g\,s)                 & $4.46 \pm 0.04$       & ARES+MOOG\\
	Metallicity $[$Fe/H$]$                  & (dex)                     & $0.256 \pm 0.048$     & ARES+MOOG \\
 	Stellar radius $R_\star$                & (\mbox{R$_{\odot}$})      & $0.87^{+0.03}_{-0.02}$& Joint fit\\
	Stellar mass $M_\star$                  & (\mbox{M$_{\odot}$})      & $0.88 \pm 0.02$       & Joint fit \\
	Rotational velocity $v \sin i_\star$    & (km\,s$^{-1}$)            & $< 1.5 \pm 1.2$       & SME \\
	Chromospheric activity index $\log R'_{\rm HK}$ & -                 & $-4.831 \pm 0.003$    & ACTIN2 \\
    Rotation period $P_{\rm rot}$           & (days)                    & $35.6 \pm 4.6$        & $\log R'_{\rm HK}$ + \citet{Mamajek2008}  \\
    Age $\tau$                              & (Gyr)                     & $5.0 \pm 2.3$         & $\log R'_{\rm HK}$ + \citet{Mamajek2008} \\
	\midrule
	\multicolumn{4}{l}{\textbf{Planet}} \\
	Period $P$                              & (days)                    & $0.777038 \pm 0.000001$ & Joint fit\\
	Full transit duration $T_{dur}$         & (hours)                   & $1.52 \pm 0.03$           & Joint fit (derived) \\
	Reference time of midtransit $t_c$     & (BJD-2457000)             & $2062.4447^{+0.0006}_{-0.0005}$    & Joint fit \\
	Radius $R_p$                            & (R$_{\oplus}$)            & $3.20^{+0.16}_{-0.11}$    & Joint fit \\
    Planet-to-star radius ratio $R_p/R_\star$ & -                         & $0.0341 \pm 0.0009$       & Joint fit (derived) \\
	Impact parameter $b$                    & -                         & $0.25^{+0.13}_{-0.15}$    & Joint fit\\  
	Inclination $i$                         & ($^{\circ}$)              & $86.4^{+2.3}_{-2.0}$      & Joint fit\\
	Eccentricity $e$                        & -                         & 0 (fixed)                 & Joint fit\\
	The argument of periastron $\omega$     & ($^{\circ}$)              & 0 (fixed)                 & Joint fit\\
	Radial velocity semi-amplitude $K$      & (ms$^{-1}$)               & $43 \pm 1$                & Joint fit\\
	Mass $M_p$                              & (M$_{\oplus}$)            & $57.2 \pm 1.6$            & Joint fit (derived) \\
	Bulk density $\rho$                     & (g\,cm$^{-3}$)            & $9.6^{+1.1}_{-1.3}$       & Joint fit (derived) \\
	Semi-major axis $a$                     & (AU)                      & $ 0.0159 \pm 0.0001$      & Joint fit (derived) \\
    System scale $a/R_{\star}$                  & -                         & $3.94^{+0.11}_{-0.12}$    & Joint fit (derived) \\
	Equilibrium temperature$^*$ $T_{\rm eq}$& (K)                       & $1871^{+30}_{-25}$        & Joint fit (derived) \\
	\bottomrule
 	\end{tabular}
	\begin{tablenotes}
	\item $^*$Equilibrium temperature is calculated assuming an albedo of zero.
	\end{tablenotes}
	\end{threeparttable}
\end{table*}

To fit the HARPS RVs, we use DACE\footnote{The DACE platform is available at \url{https://dace.unige.ch}} with a simple Keplerian model to estimate prior values for the systematic RV offset and the semi-amplitude of the RV signal $K$. We set wide, uninformative uniform priors on $K$ and the offset. We also incorporate a separate jitter term with a wide Gaussian prior, the mean of which is the log of the minimum error on the HARPS data. This term encapsulates any uncharacterised signal or noise that is perceived as white noise in the RV data, for example instrumental effects and short-scale stellar activity. As the RV data does not show any significant stellar activity by visual inspection and in the stellar activity indicators (see Figs.\,\ref{fig:rvcorr} and \ref{fig:rvperiodogram}), and does not exhibit any long-term trends, we do not perform any further detrending to it. This completes the joint fit model.

We use {\tt exoplanet} to maximise the log probability of the model. The fit values that this optimisation obtains are then used as the starting point of the {\tt PyMC3} sampler, which draws samples from the posterior using a variant of Hamiltonian Monte Carlo, the No-U-Turn Sampler (NUTS). From examination of the chains from earlier test runs of the model, we use 5 chains of 50000 steps, 1000 steps of which are discarded as burn-in. To test for non-convergence, we calculate the rank-normalised split-$\hat{R}$ statistic \citep{Vehtari2021} for each parameter. $\hat{R} \approx 1.0$ for all parameters, implying convergence. We present our best fit parameters for the TOI-332 system from this joint fit in Table \ref{tab:params}.

\section{Results and discussion}\label{resultsdiscussion}

The results of our joint fit model show that, with an orbital period of $0.777038 \pm 0.000001$\,d, TOI-332\,b is an ``Ultra-Short Period'' (USP) planet, defined as a planet with $P_{\rm orb} < 1$\,d \citep{Winn2018}. The host star, TOI-332, is a K0 dwarf with a mass of $0.88 \pm 0.02$\,M$_{\odot}$ and a radius of $0.87^{+0.03}_{-0.02}$\,R$_{\odot}$. Assuming an albedo of zero, the proximity of the planet to the star gives the planet an equilibrium temperature of $1871^{+30}_{-26}$\,K; it is highly irradiated, receiving approximately 2400 times the instellation of the Earth per unit area. 

TOI-332\,b has a mass of $57.2 \pm 1.6$\,M$_{\oplus}$, more than half the mass of Saturn, yet a radius of $3.20^{+0.16}_{-0.11}$\,R$_{\oplus}$, smaller than that of Neptune. With a density of $9.6^{+1.1}_{-1.3}$\,g\,cm$^{-3}$, it is one of the densest planets of those with the size of Neptune or greater found thus far (Fig.\,\ref{fig:mrdiagram}). 
These parameters place TOI-332\,b deep in the Neptunian desert (Fig.\,\ref{fig:nepdesert}). 

Taking all of this into account, TOI-332\,b is a very interesting addition to our current Neptunian desert discoveries and a case study to test planet formation theory.

\subsection{Interior structure}\label{structure}

\begin{figure*}
    \centering
    \includegraphics[width=0.8\textwidth]{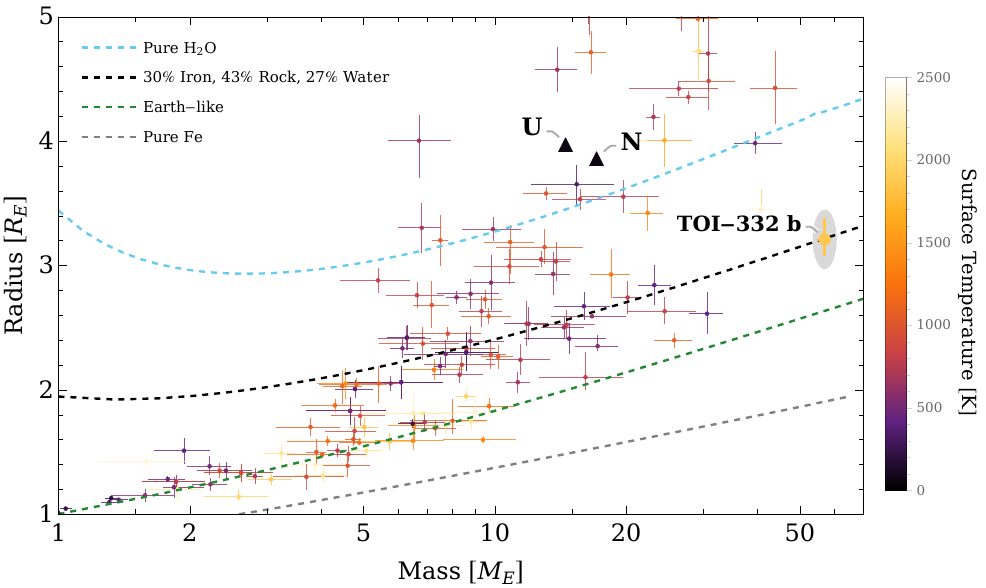}
    \caption{Mass-radius diagram of the exoplanets in the Otegi catalog \citep{Otegi2020}. The color of the planets indicates their equilibrium temperature. The dashed blue, green, and gray line show the mass-radius relation for a pure water, an Earth-like, and a pure iron composition at TOI-332 b's equilibrium temperature (1869.4 K), respectively. Uranus and Neptune are shown as black triangles.}
    \label{fig:mrdiagram}
\end{figure*}

As seen in Fig.\,\ref{fig:mrdiagram}, TOI-332\,b  occupies a unique and unpopulated spot in the mass-radius (M-R) diagram. Its mass and radius suggest a composition that is dominated by refractory materials, potentially more similar to that of terrestrial planets.

To put limits on the possible composition of TOI-332\,b, we use a layered interior model similar to those used in \citet{Dorn2017} and \citet{Armstrong2020}. This model consists of up to four layers including an iron core, a silicate mantle, a water layer, and a H-He atmosphere. For these layers, we solve the standard structure equations to estimate the possible ranges of H-He mass fractions. We note, however, that for such high mass planets, layers might not be as distinct as assumed here \citep[e.g.,][]{Helled2017,Bodenheimer2018}. Overall, the planet is found to consist of 30\,per\,cent iron core, 43\,per\,cent rock mantle, 27\,per\,cent water, and a negligible H-He envelope.

To constrain the H-He mass, we investigate the extreme situation of a planet without water and compare it with a planet where the water abundance is allowed to vary freely. For these cases, we construct structure models that reproduce the measured  mass and radius of TOI-332 b. Moreover, we assume host star elemental abundances.

We find that even if TOI-332\,b had no water, the H-He mass fraction would be only $1.8^{+0.6}_{-0.5}~\%$. In the water-containing model, the H-He mass fraction is $ \log(M_\mathrm{atm}/M_\mathrm{p}) = -6.7\pm 3.2$, well below 0.1\,per\,cent. We can therefore conclude that, unless the planet is devoid of water, the atmospheric mass of TOI-332 b is very small. 

Typically, planets with comparable masses to TOI-332\,b are expected to be H-He dominant in composition, and terrestrial planets are not expected to reach several tens of Earth masses. Planetary embryos are expected to accrete only a few to $\sim$20 Earth masses of heavy elements before the onset of rapid gas accretion \citep[e.g., ][]{Pollack1996,Lambrechts2014,Piso2015}, resulting in a large envelope. However, with such a large core mass and little envelope, the existence of TOI-332\,b requires further explanation, perhaps having lost an initial envelope, or having managed to avoid core-accretion. We explore several scenarios below.

\subsection{Co-orbital bodies}\label{coorbital}

To try and explain the apparently excessive core mass of TOI-332\,b, we first tested an alternative co-orbital configuration of two planets which may mimic the appearance of a single, more massive planet, and compare its evidence against the current one-planet scenario. 

Co-orbital exoplanets (pairs of planets trapped in 1:1 resonances) are dynamically stable under very soft conditions \citep{Laughlin2002}, and several formation mechanisms have already been proposed for these configurations \citep[see e.g., ][]{Beauge2007,Namouni2017,Leleu2019}. However, no co-orbital exoplanets have yet been found despite several efforts \citep[e.g., ][]{Janson2013,Hippke2015,Ford2007,Madhusudhan2009,LilloBox2018a,LilloBox2018b}, although different candidates have already been proposed \citep[e.g., ][]{LilloBox2014,Boyajian2016,LilloBox2020}. 

In the particular case of TOI-332\,b, we explore the scenario where this planet is actually a pair of planets in 1:1 resonance where the lighter planet transits the host star while the more massive component does not. Assuming a low eccentricity scenario, we can approximate the sum of two Keplerians with the same periodicity as a single Keplerian. We might attribute the mass of the more massive (non-transiting) component to the only component that we see transiting the host star. This will imprint specific features in the radial velocity data that are testable through available techniques. In particular, in order to test this scenario, we apply the technique described in \citet{Leleu2017} (a generalisation of the technique proposed by \citet{Ford2006}), which combines the transit and radial velocity information to infer time lags between the time of transit and the radial velocity phase. This technique is based on the modelling of the radial velocity data assuming the time of conjunction and period derived from the transit modelling. In this case, we assume the value $t_c=2459062.444852864292$ and and $P=0.77703814$~days, obtained only by modelling the photometry from {\it TESS} and LCOGT. We apply Equation 18 in \citet{Leleu2017}, which includes the RV semi-amplitude $K$, the orbital configuration parameters ($c$ and $d$), and the additional parameter $\alpha$, a measure of the mass imbalance between the Lagrangian regions L$_{\rm 4}$ and L$_{\rm 5}$. 

Here we test four different models: two assuming only one planet in the orbit (one assuming a circular orbit, ``1p(c)'', the other including the possibility of a slightly eccentric orbit, ``1p''), and two including the co-orbital scenario (again, one assuming a circular orbit, ``1p(c)T'', and the other leaving the eccentricity as a free parameter, ``1pT''). We use the implementation of \citet{GoodmanWeare}'s affine invariant Markov chain Monte Carlo (MCMC) ensemble sampler \texttt{emcee} \citep{ForemanMackey2013} to sample the posterior probability distribution of each of these parameters. The MCMC chains are subsequently used to estimate the Bayesian evidence ($\ln{\mathcal{Z}_i}$) of the models using the \texttt{perrakis} implementation \citep{Diaz2016}. 

The results of our analysis show the model with just one planet in circular orbit as the most favourable model based on the current dataset ($\Delta \ln{\mathcal{Z}} > +7$ against the other more complex models). Consequently we can conclude that the current dataset does not support the presence of an additional co-orbital planet, hence confirming that all the mass at this periodicity is accumulated into the transiting body. For the simpler co-orbital model with circular orbit, we obtain $\alpha=-0.031^{+0.032}_{-0.031}$, hence compatible with zero (i.e., no mass imbalance between L4 and L5 and so potentially no co-orbitals) at the 1$\sigma$ level. 

\begin{figure*}
    \centering
    \includegraphics[width=\textwidth]{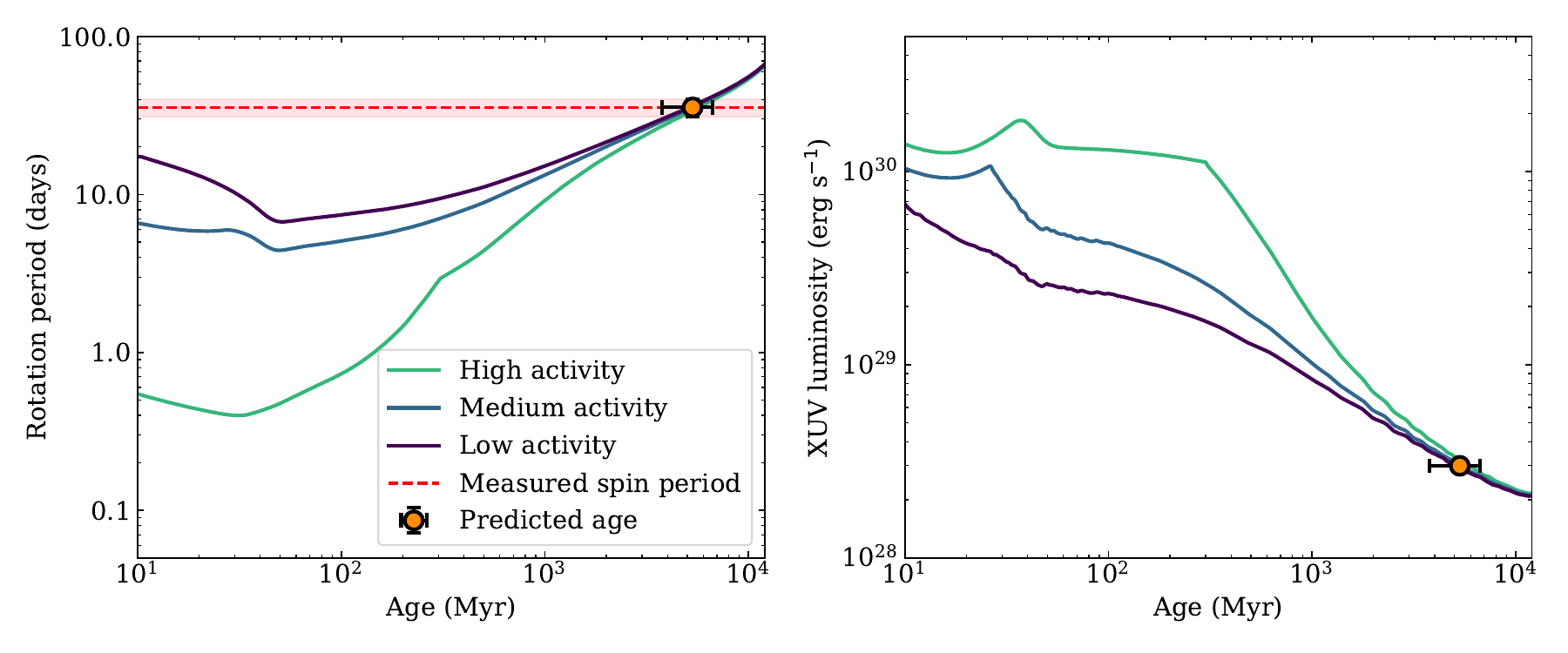}
    \caption{\textbf{Left panel:} plot of rotation period against age showing rotational evolution models by \citet{Johnstone2021}, with high, medium, and low activity tracks for a $0.9\,$M$_\odot$ star. Its measured rotation period is shown as a dashed red line, with the uncertainty as a shaded region. The age estimated with gyrochronology is plotted as an orange circle.
    \textbf{Right panel:} plot of XUV luminosity against age showing the corresponding XUV evolution tracks to the models on the left panel, as well as the predicted XUV luminosity based on its rotation period. The models were calculated using the methods described in Section\,\ref{photoevap}.}
    \label{fig:jorge-star-evo}
\end{figure*}

\begin{figure*}
    \centering
    \includegraphics[width=\textwidth]{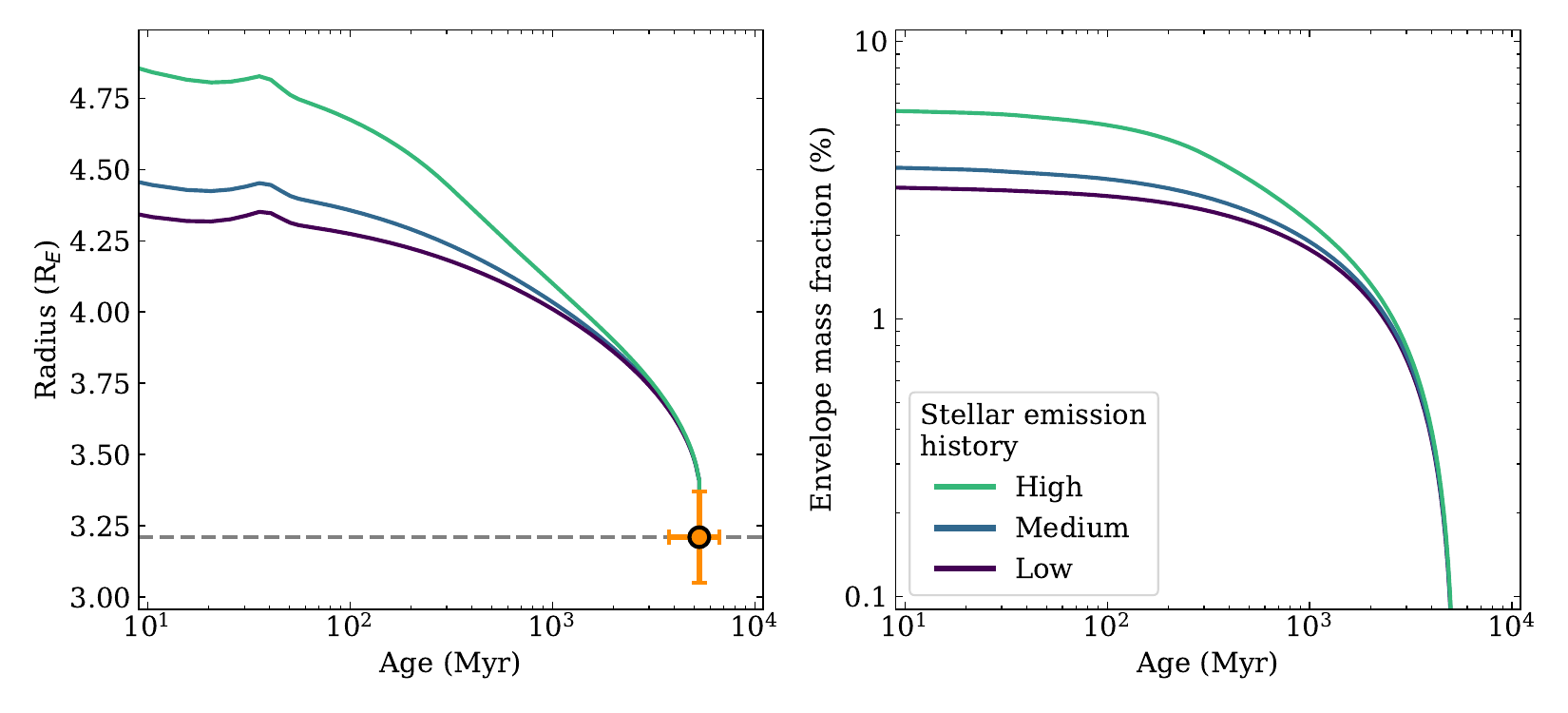}
    \caption{\textbf{Left panel}: plot of planet radius against age showing the evolution of the radius of TOI-332\,b using the three stellar XUV emission histories described in Section\,\ref{photoevap}.
    \textbf{Right panel:} plot of envelope mass fraction against age showing the evolution of the past envelope mass of TOI-332\,b following the left panel.}
    \label{fig:jorge-planet-evo}
\end{figure*}

\subsection{Evolution under XUV-driven escape}\label{photoevap}

We then test whether a TOI-332\,b-like planet could be reproduced by stripping an initial accreted envelope through X-ray and extreme-ultraviolet (EUV; together, XUV) driven escape. 

We performed simulations on the evaporation history of TOI-332\,b by taking into account the range of possible XUV emission histories of the star motivated by its stellar parameters. We fitted the star's inferred spin period of $35.6 \pm 4.6$\,days with the rotational evolution models of \citet{Johnstone2021} and estimated a gyrochronological age of $5.3^{+1.5}_{-1.4}$\,Gyr.

By field age, the initial spread in stellar rotation periods have largely converged to a single track, leaving their histories degenerate, i.e., we cannot tell which history TOI-332 followed. We thus considered three spin histories in order to sample the diversity of possible X-ray activity pasts experienced by the planet: the \textit{low}, \textit{medium}, and \textit{high} activity scenarios, which represent the model's 5th, 50th, and 95th percentiles in the distribution of rotation periods at any given age. These rotational histories are shown in Fig.\,\ref{fig:jorge-star-evo} (left hand panel), together with the star's current place along these tracks.
The corresponding XUV luminosity tracks for these scenarios are shown in Fig.\,\ref{fig:jorge-star-evo} (right hand panel).

We then simulated the evaporation history of TOI-332\,b using the \textit{photoevolver} code\footnote{The evaporation evolution code is available on GitHub at \url{https://github.com/jorgefz/photoevolver}} \citep{Fernandez2023}. For this analysis, we adopted the full hydrodynamic model of \citet{Kubyshkina2018} (and the interpolation routine of \citet{KubyshkinaFossati2021}) to calculate the mass loss rates.

Taking the results of the interior structure characterisation, we assumed that the planet is currently a bare core with no gaseous envelope. We can thus estimate an upper limit on the initial envelope mass fraction assuming that it has just finished evaporating. Over this upper limit, the envelope would fail to evaporate in the lifetime of the planet, and this would be inconsistent with the planet's current structure.

We achieved this by adding a tiny amount of gas to the planet, equivalent to 0.01\,per\,cent of its total mass (such that it is completely evaporated within one simulation time step), and evolved this tenuous atmosphere backwards in time to the age of 10\,Myr. We repeated this process using each of the three XUV emission scenarios, and plot the results in Fig.\,\ref{fig:jorge-planet-evo}.

We find that the possible evaporation histories for TOI-332\,b based on these scenarios lead to a narrow range of upper limits on the initial envelope mass fraction, between 3 and 6\,per\,cent. We thus find that TOI-332\,b starting out as a Jupiter-sized planet is inconsistent with photoevaporation as the only mechanism for mass loss.

\subsection{Other formation scenarios}

It is clear that if TOI-332\,b originally had a Jupiter-like envelope as we would expect for a core of this size, photoevaporation could not have been the sole mechanism responsible for the removal of most of its atmosphere. So we can theorise other scenarios that could have caused this. An initially large envelope may have been removed by high-eccentricity migration and subsequent tidal thermalisation \citep[e.g.][]{Ivanov2004,Vick2018,Wu2018,Vick2019}. Alternatively, the atypical composition of TOI-332\,b could be the result of a giant impact between two gas giants followed by efficient removal of the gaseous atmosphere \citep[e.g.][]{Liu2015,Emsenhuber2021,Ogihara2021}. Finally, runaway accretion could have just been avoided entirely by, for example, gap opening in the protoplanetary disk \citep[e.g.][]{Crida2006,Duffell2013,Lee2019}. 

However, we do not think it is currently possible to say which, if any, of these formation scenarios created TOI-332\,b, though future observations may aid us in this.

\subsection{Orbital decay rate}\label{orbdecay}

As TOI-332\,b is both unusually massive and close to its host star, it may be one of the most well-placed non-gas giant planets for an orbital decay rate study. 

We follow the method outlined in \citet{Jackson2023} to calculate the orbital decay rate, $dP/dt$. In short, we use the following equation \citep{Goldreich1966,Ogilvie2014}:

\begin{equation}
    \frac{dP}{dt}=-\frac{27\pi}{2Q'_*}\left(\frac{M_p}{M_\star}\right)\left(\frac{R_*}{a}\right)^5
\end{equation}

\noindent where $Q'_*=3Q_*/2k_2$. $Q_*$ is the tidal quality factor, $Q'_*$ is the reduced tidal quality factor, and $k_2$ is the dimensionless quadrupolar Love number.

We do not consider the dynamical tide within the convective zone, only the equilibrium tide, as the orbital period of the planet is much less than twice the stellar rotation period. This allows us to calculate $Q'_*$ as defined in \citet{Strugarek2017}, which requires a value for the depth of the convective zone\footnote{We estimate the radius and mass of the stellar core using \url{http://www.astro.wisc.edu/~townsend/static.php?ref=ez-web}.}. 

We find $Q'_* = 8 \times 10^6$, resulting in $dP/dt = -1.05 \times 10^{-12}$. We use the decay rate together with the period to estimate the length of time it would take for the orbit to decay completely (i.e. reaching a period of zero) as $2.0$\,Gyr. We note that this is a likely upper-estimate of the decay timescale, as it assumes a constant rate of decay, ignoring any effects that may alter this (e.g. stellar wind and stellar evolution). Our method also does not take into account the structure of the planet, which some other more complex treatments of tidal effects do \citep[e.g.][]{Henning2014,Clausen2015,Brasser2019}, but this is beyond the scope of this paper.

We can also estimate boundaries on this decay timescale by assuming upper and lower limits on $Q'_*$ of $10^5$ and $10^8$, resulting in timescales of 0.25 and 250\,Gyr respectively. Even the shortest decay timescale of 0.25\,Gyr is several magnitudes longer than the current decay timescale estimate for WASP-12\,b of $3.16 \pm 0.10$\,Myr, to date the only planet we have confidently detected an orbital decay for \citep{Maciejewski2016,Patra2017,Patra2020,Yee2020,Wong2022}. Thus we conclude that measuring the orbital decay of TOI-332\,b is not going to be possible over a realistic span of time.

\subsection{Future observation prospects}

TOI-332\,b is undoubtedly an unusual and unique planet, and further observations will be needed to deduce more about its formation and evolutionary history and its current composition.

The Rossiter-McLaughlin (RM) effect allows us to measure the sky-projected obliquity of a system, and is important for constraining formation scenarios: disk-migration is expected to conserve alignment between the angular momentum of a disk and planetary orbits, but misalignment could imply, for example, planet-planet/planet-star scattering, high-eccentricity migration,  or tidal disruption. If TOI-332\,b lacks an atmosphere due to reduced gas accretion through gap opening, it should align to the stellar spin axis, but if it is misaligned, it might imply a more violent history has removed an initial envelope - though at such a short period, there is the possibility that tides might cause realignment even if the orbit and stellar spin axis began misaligned. We can predict a RM semi-amplitude of approximately 2.1\,m\,s$^{-1}$ \citep{Triaud2018}; though this signal is small, smaller RM amplitudes have been measured \citep[e.g.][]{Winn2010,Bourrier2014} and are obtainable with high-precision spectrographs like HARPS/HARPS-N and ESPRESSO, and methods such as the Rossiter-McLaughlin effect Revolutions (RMR) technique \citep{Bourrier2021}.

There is evidence from other USP planet discoveries that they often have companions with periods out to 50 days \citep{Sanchis-Ojeda2014}, and for a system like TOI-332\ with $a/R_{\star} < 5$ they'd be expected to have a minimum mutual inclination of $5 - 10\deg$ \citep{Dai2018}. We find no evidence of a companion in our current data, photometric or spectroscopic; further long-term monitoring of this system would be needed to discover or rule out a companion, and this may also help narrow down formation scenarios for this system.

Characterising potentially unusual atmospheres and surfaces of highly-irradiated rocky worlds is an exciting prospect. While the predicted atmospheric mass fraction of TOI-332\,b is small, the high equilibrium temperature of TOI-332\,b may lead to evaporation of volatiles and formation of a secondary atmosphere that could contain core materials. The composition of such an atmosphere could be determined with JWST. Additionally, JWST could be used to obtain a phase curve of TOI-332\,b, which would constrain its dayside and nightside temperatures and any phase offset, its Bond albedo, and heat recirculation efficiency. With little atmosphere we would expect high temperature contrast and poor recirculation, and may be able to distinguish between different surface composition scenarios. 

\section{Conclusion}\label{conclusion}

We present here the discovery and characterisation of a new planet in the TOI-332 system. We use photometry from two {\it TESS} sectors at 30\,min cadence and one sector at 2\,min cadence, plus six LCOGT transit events. There is further photometry from PEST and WASP-South, but this was not included in the final fit due to the ambiguity of the transit detections. The photometric data were modelled jointly with 16 RV data points from the HARPS spectrograph. Multiple sources of high-resolution imaging confirm that the star is single with no unresolved companions. 

The planet TOI-332\,b is on an ultra-short period of 0.78 days, with a radius smaller than Neptune but an anomalously large mass of more than half that of Saturn, making it one of the densest known Neptune-sized planets discovered thus far. It is located deep within the Neptunian desert, and is one of only a handful of planets that have been found there, being one of even fewer to have a precise mass determination. Using a four layer model consisting of an iron core, silicate mantle, water, and a H-He envelope, interior structure characterisation determines that it likely possesses a negligible H-He envelope. 

This unusual planet tests what we currently understand about planet formation; how such a giant core exists without a gaseous envelope remains an unanswered question. We determine that photoevaporation would be insufficient on its own in removing a Jupiter-like envelope, and we instead posit high-eccentricity migration or giant impacts as possible mechanisms for stripping the initial envelope from TOI-332\,b. Alternatively, a mechanism like disc-gap opening could have led it to avoid gas accretion in the first instance. Further observations are needed to potentially disentangle TOI-332\,b`s formation history and current characteristics.

\section*{Acknowledgements}

We thank the anonymous referee whose constructive comments have contributed to the quality of this paper.

AO, JFF, and FH are funded by STFC studentships.

DJA is supported by UKRI through the STFC (ST/R00384X/1) and EPSRC (EP/X027562/1).

HK, RH, and HPO carried out this work within the framework of the National Centre of Competence in Research PlanetS (NCCR PlanetS) supported by the Swiss National Science Foundation under grants 51NF40\_182901 and 51NF40\_205606. The authors acknowledge the financial support of the SNSF.

KAC acknowledges support from the TESS mission via subaward s3449 from MIT.

EDM acknowledges the support from FCT through Investigador FCT contract nr. 2021.01294.CEECIND.

MF gratefully acknowledges the support of the  Swedish National Space Agency (DNR 65/19, 174/18, 177/19,  2020-00104).

DGJ acknowledges the support of the Department for Economy (DfE).

JL-B acknowledges financial support received from "la Caixa" Foundation (ID 100010434) and from the European Unions Horizon 2020 research and innovation programme under the Marie Slodowska-Curie grant agreement No 847648, with fellowship code LCF/BQ/PI20/11760023. This research has also been partly funded by the Spanish State Research Agency (AEI) Project No.PID2019-107061GB-C61.

NCS is funded/co-funded by the European Union (ERC, FIERCE, 101052347). Views and opinions expressed are however those of the author(s) only and do not necessarily reflect those of the European Union or the European Research Council. Neither the European Union nor the granting authority can be held responsible for them. This work was supported by FCT - Fundação para a Ciência e a Tecnologia through national funds and by FEDER through COMPETE2020 - Programa Operacional Competitividade e Internacionalização by these grants: UIDB/04434/2020; UIDP/04434/2020.

SGS acknowledges the support from FCT through Investigador FCT contract nr. CEECIND/00826/2018 and POPH/FSE (EC).

ODSD is supported in the form of work contract (DL 57/2016/CP1364/CT0004) funded by national funds through Fundação para a Ciência e Tecnologia (FCT). This work was also supported by FCT through national funds by the following grants: 2022.06962.PTDC.

CD acknowledges support from the Swiss National Science Foundation under grant PZ00P2\_174028. 

This project has received funding from the European Research Council (ERC) under the European Union’s Horizon 2020 research and innovation programme (grant agreement SCORE No 851555)

PJW acknowledges support from STFC under consolidated grant ST/T000406/1. 


This research has made use of the NASA Exoplanet Archive, which is operated by the California Institute of Technology, under contract with the National Aeronautics and Space Administration under the Exoplanet Exploration Program.

This research made use of {\tt exoplanet} \citep{exoplanet:exoplanet} and its dependencies \citep{exoplanet:agol20,
exoplanet:arviz, exoplanet:astropy13, exoplanet:astropy18, exoplanet:kipping13,
exoplanet:luger18, exoplanet:pymc3, exoplanet:theano}.

This publication makes use of The Data \& Analysis Center for Exoplanets (DACE), which is a facility based at the University of Geneva (CH) dedicated to extrasolar planets data visualisation, exchange and analysis. DACE is a platform of the Swiss National Centre of Competence in Research (NCCR) PlanetS, federating the Swiss expertise in Exoplanet research. The DACE platform is available at \url{https://dace.unige.ch}.

This paper made use of data collected by the TESS mission and are publicly available from the Mikulski Archive for Space Telescopes (MAST) operated by the Space Telescope Science Institute (STScI). Funding for the TESS mission is provided by NASA’s Science Mission Directorate. We acknowledge the use of public TESS data from pipelines at the TESS Science Office and at the TESS Science Processing Operations Center. Resources supporting this work were provided by the NASA High-End Computing (HEC) Program through the NASA Advanced Supercomputing (NAS) Division at Ames Research Center for the production of the SPOC data products.

This research has made use of the Exoplanet Follow-up Observation Program (ExoFOP; DOI: 10.26134/ExoFOP5) website, which is operated by the California Institute of Technology, under contract with the National Aeronautics and Space Administration under the Exoplanet Exploration Program.

This work makes use of observations from the LCOGT network. Part of the LCOGT telescope time was granted by NOIRLab through the Mid-Scale Innovations Program (MSIP). MSIP is funded by NSF.

Some of the observations in this paper made use of the High-Resolution Imaging instrument Zorro and were obtained under Gemini LLP Proposal Number: GN/S-2021A-LP-105. Zorro was funded by the NASA Exoplanet Exploration Program and built at the NASA Ames Research Center by Steve B. Howell, Nic Scott, Elliott P. Horch, and Emmett Quigley. Zorro was mounted on the Gemini South telescope of the international Gemini Observatory, a program of NSF’s OIR Lab, which is managed by the Association of Universities for Research in Astronomy (AURA) under a cooperative agreement with the National Science Foundation. on behalf of the Gemini partnership: the National Science Foundation (United States), National Research Council (Canada), Agencia Nacional de Investigación y Desarrollo (Chile), Ministerio de Ciencia, Tecnología e Innovación (Argentina), Ministério da Ciência, Tecnologia, Inovações e Comunicações (Brazil), and Korea Astronomy and Space Science Institute (Republic of Korea).

This work made use of \texttt{tpfplotter} by J. Lillo-Box (publicly available in \url{www.github.com/jlillo/tpfplotter}), which also made use of the python packages \texttt{astropy}, \texttt{lightkurve}, \texttt{matplotlib} and \texttt{numpy}.

\section*{Data Availability}

The {\it TESS} data are available from the Mikulski Archive for Space Telescopes (MAST), at \url{https://heasarc.gsfc.nasa.gov/docs/tess/data-access.html}. The other photometry from LCOGT and PEST, and the high-resolution imaging data, are available for public download from the ExoFOP-TESS archive at \url{https://exofop.ipac.caltech.edu/tess/target.php?id=139285832}. The full HARPS RV data products are publicly available from the ESO archive, at \url{http://archive.eso.org/wdb/wdb/adp/phase3_main/form}. The model code underlying this article will be shared on reasonable request to the corresponding author. The MCMC chains are available from Zenodo, at \url{https://doi.org/10.5281/zenodo.8199962}.


\bibliographystyle{mnras}
\bibliography{paper}



\appendix

\section{Photometry}

\begin{figure*}
    \centering
    \includegraphics[width=0.8\textwidth]{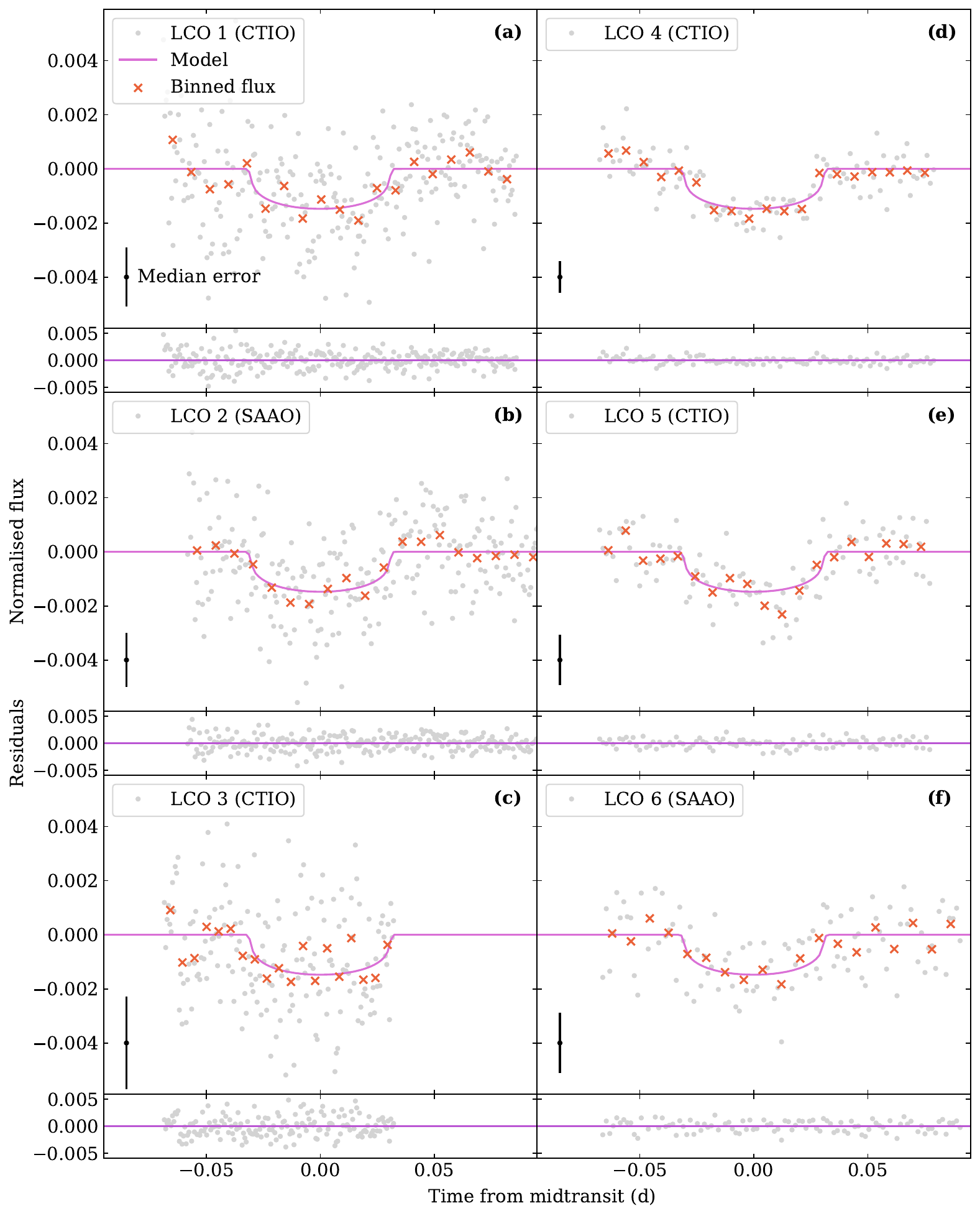}
    \caption{Photometric data from LCOGT. For each, the flux (grey circles), binned flux (red crosses), median error on the flux (one standard deviation, black error bar, bottom left) and best fit model (solid line) are shown. Residuals after the best fit model is subtracted are shown in the bottom panels. Data were captured at two different telescopes in the Global Network, the Cerro Tololo Inter-American Observatory (CTIO) and the South African Astronomical Observatory (SAAO), on the nights of: \\
    \textbf{(a)} 1 June 2019 at CTIO in Sloan $i'$ band (``LCO 1''); \\
    \textbf{(b)} 10 July 2019 at SAAO in Sloan $i'$ band (``LCO 2''); \\
    \textbf{(c)} 27 July 2019 at CTIO Pan-STARRS $z$-short band (``LCO 3''); \\
    \textbf{(d)} 10 Aug 2019 at CTIO in Sloan $g'$ band (``LCO 4''); \\
    \textbf{(e)} 10 Aug 2019 at CTIO in Pan-STARRS $z$-short band (``LCO 5''); \\
    and \textbf{(f)}, 24 Aug 2020 at SAAO in Pan-STARRS $z$-short band (``LCO 6'').}
    \label{fig:lco}
\end{figure*}

\begin{figure}
    \centering
    \includegraphics[width=\columnwidth]{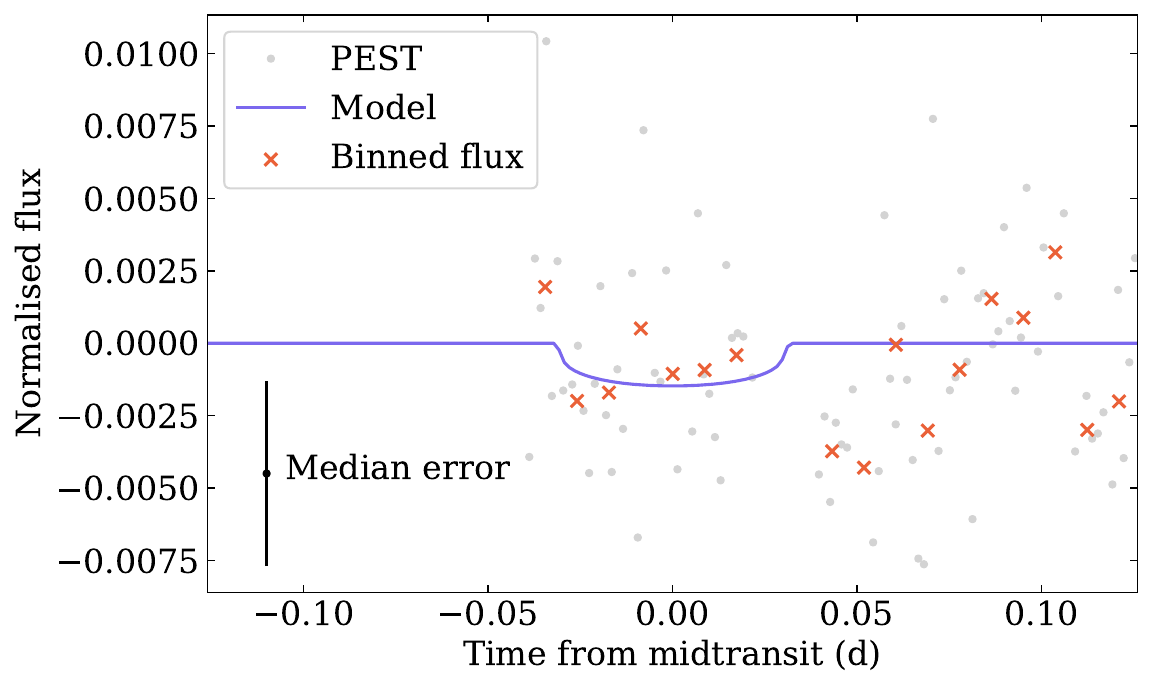}
    \caption{Photometric data from PEST (a single transit event), where the flux (grey circles), binned flux (red crosses), and median error on the flux (one standard deviation, black error bar, bottom left) are shown. This data was not included in our joint fit model (see Section\,\ref{pest}), but we overplot the best fit model with a solid line.}
    \label{fig:pest}
\end{figure}

\begin{figure}
    \centering
    \includegraphics[width=0.9\columnwidth]{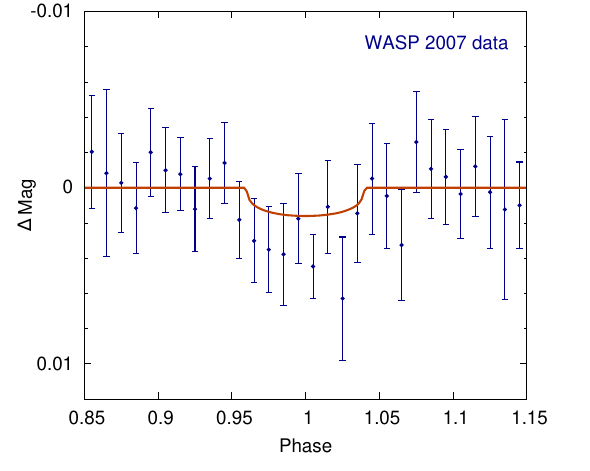}
    \caption{Photometric data from WASP (2007 data only), phase-folded on the transit ephemeris. This data was not included in our joint fit model (see Section\,\ref{wasp}), but we overplot the best fit model with a solid line.}
    \label{fig:wasp}
\end{figure}

\begin{figure}
  \includegraphics[width=\columnwidth]{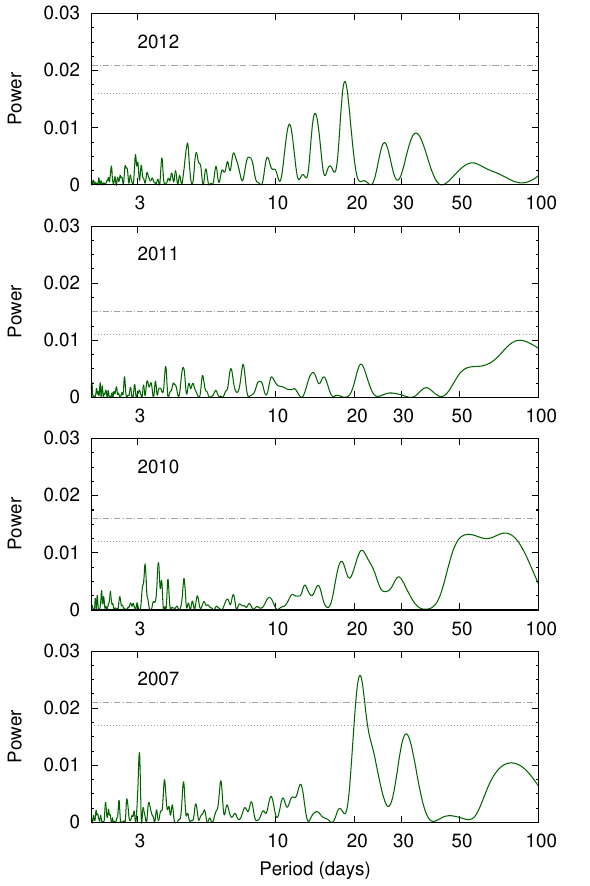}
  \caption{Periodograms of the WASP-South data for TOI-332 from four different years of data.  The horizontal lines mark the estimated 10\,per\,cent- and 1\,per\,cent-likelihood false-alarm levels. A significant periodicity at $20.9 \pm 1.0$\,d was seen in 2007.}
\label{fig:wasp_rotmod}
\end{figure}

The full set of six transits from LCOGT described in Section\,\ref{lcogt} are shown in Fig.\,\ref{fig:lco}. The PEST transit described in Section\,\ref{pest} is shown in Fig.\,\ref{fig:pest}. Described in Section\,\ref{wasp}, the phase folded WASP data  are shown in Fig.\,\ref{fig:wasp}, and the periodograms from each of the four different years of WASP data are shown in Fig.\,\ref{fig:wasp_rotmod}. 

\section{Spectroscopy}

The HARPS RV data (described in Section\,\ref{harps}) are presented in Table \ref{tab:rvs}. 

\begin{table*}
    \small
	\caption{HARPS radial velocities.}
	\label{tab:rvs}
	\begin{threeparttable}
	\begin{tabular}{lllllll}
	\toprule
	Time                & RV            & $\sigma_\textrm{RV}$  & FWHM          & Bisector      & Contrast  \\
	(RJD)               & ($ms^{-1}$)   & ($ms^{-1}$)           & ($ms^{-1}$)   & ($ms^{-1}$)   &           \\
	\midrule
	59544.57860093983   & -6694.929826  & 2.367355  & 6392.280263   & 26.740961 & 46.535725& \\
    59545.55694768019   & -6649.202052	& 2.955035	& 6388.488003	& 25.186455	& 46.576300& \\
	59547.619848280214  & -6715.203341	& 3.033317	& 6385.913640	& 18.836766	& 46.524531& \\
	59548.59424903989   & -6657.213721	& 2.254944	& 6393.227848	& 23.858339	& 46.540726& \\
	59549.57656604005   & -6671.028545	& 2.193774	& 6382.831911	& 13.896637	& 46.601883& \\
	59550.5969126299    & -6731.050809	& 3.426731	& 6391.442681	& 32.262988	& 46.546534& \\
	59573.55401741015   & -6649.345311	& 2.595591	& 6395.329891	& 10.285248	& 46.332886& \\
	59574.5364167802    & -6692.040443	& 3.867689	& 6385.126449	& 19.361261	& 45.940802& \\
	59578.54254523991   & -6732.742673	& 4.000978	& 6394.998608	& 29.815993	& 46.371621& \\
	59580.54742026003   & -6650.166676	& 3.263565	& 6384.294121	& 3.673344	& 46.562577& \\
	59581.54694268014   & -6711.154291	& 5.927420	& 6393.494683	& 34.206176	& 46.592209& \\
	59582.541859869845  & -6732.443878	& 2.576812	& 6378.507589	& 10.085069	& 46.563025& \\
	59726.90363577986   & -6723.985683	& 2.841378	& 6385.428239	& 13.959675	& 46.604519& \\
	59727.8527730098    & -6731.586132	& 2.598389	& 6379.880003	& 21.284249	& 46.570771& \\
	59728.80604236014   & -6672.012893	& 3.704808	& 6385.685635	& 36.586262	& 46.578561& \\
	59728.92009065999   & -6655.110375	& 6.944427	& 6370.694975	& 26.775309	& 46.895611& \\
	\bottomrule
	\end{tabular}
    \begin{tablenotes}
    \item The full HARPS data products can be found on ExoFOP-TESS at \url{https://exofop.ipac.caltech.edu/tess/target.php?id=139285832}
    \end{tablenotes}
    \end{threeparttable}
\end{table*}

\section{Spectroscopic analysis and chemical abundances}

\begin{figure}
    \centering
    \includegraphics[width=0.6\linewidth,trim=70 80 80 80,clip,angle=90]{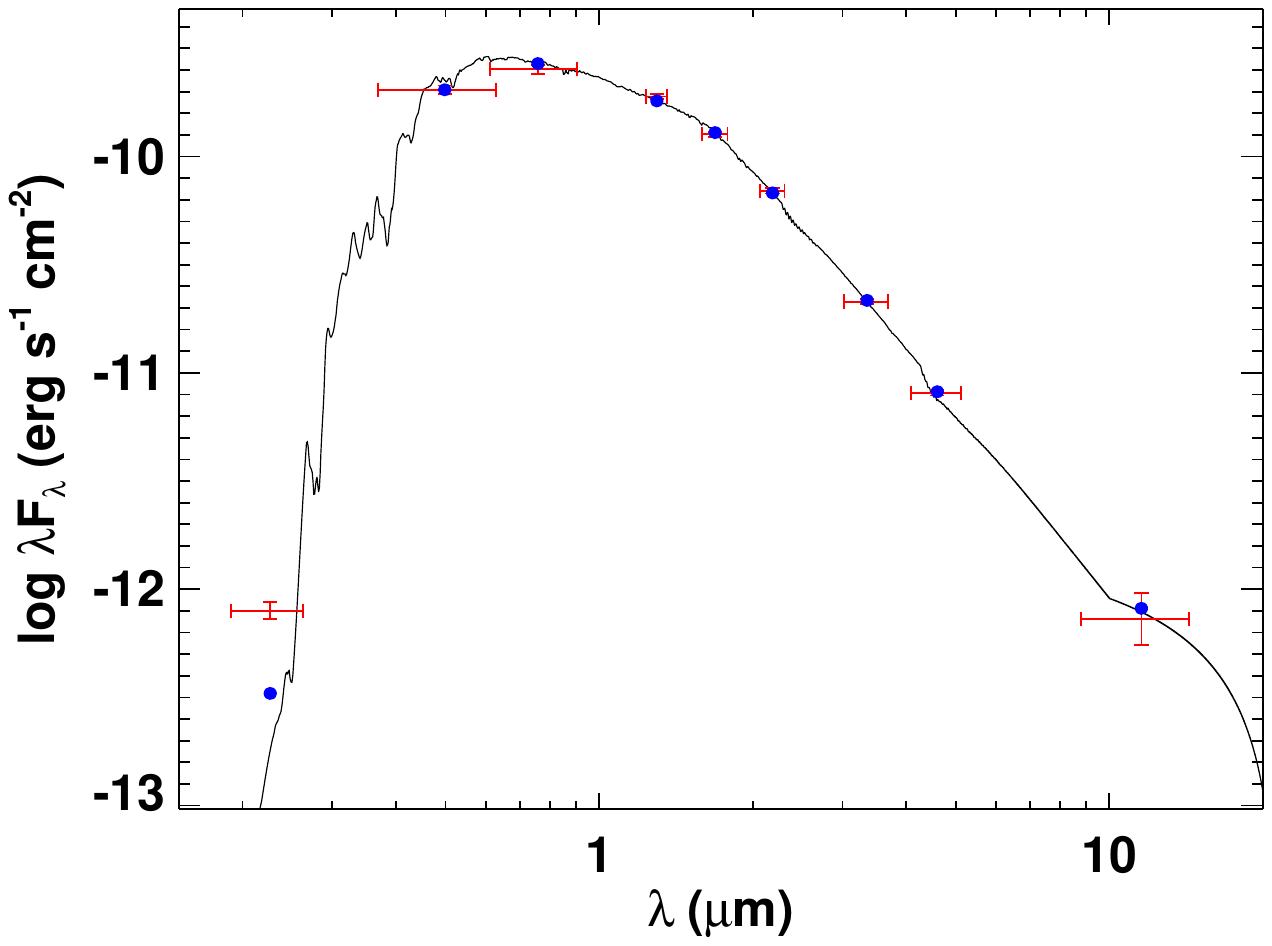}
    \caption{Spectral energy distribution of TOI-332. Red symbols represent the observed photometric measurements, where the horizontal bars represent the effective width of the passband. Blue symbols are the model fluxes from the best-fit Kurucz atmosphere model (black).}  
    \label{fig:sed}
\end{figure}

Stellar abundances determined by the methods outlined in Section\,\ref{stellar} are presented in Table\,\ref{tab:abundances}. The spectral energy distribution used for determining stellar parameters in Section\,\ref{stellar} is shown in Fig.\,\ref{fig:sed}. 

\begin{table}
    \small
    \caption{Stellar abundances determined by the methods outlined in Section\,\ref{stellar}.}
    \centering
    \begin{tabular}{ll}
    \toprule
    \textbf{Chemical abundances}     & \textbf{Value (dex)} \\
    \midrule
    $[$Ca/H$]$   & $0.22 \pm 0.06$ \\
    $[$Na/H$]$   & $0.37 \pm 0.07$ \\
    $[$Mg/H$]$   & $0.26 \pm 0.06$ \\
    $[$Al/H$]$   & $0.34 \pm 0.06$ \\
    $[$Si/H$]$   & $0.24 \pm 0.04$ \\
    $[$Ti/H$]$   & $0.33 \pm 0.06$ \\
    $[$Ni/H$]$   & $0.27 \pm 0.04$ \\
    $[$O/H$]$    & $0.23 \pm 0.15$ \\
    $[$C/H$]$    & $0.22 \pm 0.04$ \\
    $[$Cu/H$]$   & $0.41 \pm 0.09$ \\
    $[$Zn/H$]$   & $0.25 \pm 0.05$ \\
    $[$Sr/H$]$   & $0.33 \pm 0.17$ \\
    $[$Y/H$]$    & $0.18 \pm 0.11$ \\
    $[$Zr/H$]$   & $0.18 \pm 0.11$ \\
    $[$Ba/H$]$   & $0.10 \pm 0.09$ \\
    $[$Ce/H$]$   & $0.20 \pm 0.15$ \\
    $[$Nd/H$]$   & $0.27 \pm 0.08$ \\
    \bottomrule
    \end{tabular}
    \label{tab:abundances}
\end{table}

\section{The joint fit}

The full set of priors and fit values from our joint fit model described in Section\,\ref{jointfit} are presented in Table \ref{tab:jointfit}. 

\begin{table*}
    \small
    \caption{Prior distributions used in our joint fit model, fully described in Section\,\ref{jointfit}, and the fit values resulting from the model. The priors are created using distributions in \textsc{PyMC3} with the relevant inputs to each distribution described in the table footer. The fit values are given as the median values of the samples, and the uncertainties as the 16th and 84th percentiles. Further (derived) system parameters can be found in Table \ref{tab:params}.}
	\label{tab:jointfit}
	\begin{threeparttable}
	\begin{tabular}{l p{1.8cm} ll}
	\toprule
	\textbf{Parameter} & \textbf{(unit)} & \textbf{Prior Distribution} & \textbf{Fit value}  \\
	\midrule
	\multicolumn{4}{l}{\textbf{Planet}} \\
	Period $P$                      & (days)               & $\mathcal{U}(0.767, 0.787)$       & $0.777038 \pm 0.000001$ \\
	Reference time of midtransit $t_c$                 & (BJD-2457000)        & $\mathcal{U}(2062.4439, 2062.4459)$   & $2062.4447^{+0.0006}_{-0.0005}$ \\
	$\log{(R_p)}$                   & (\mbox{R$_{\odot}$}) & $\mathcal{U}(-4.6863, -2.6863)^*$          & $-3.53^{+0.05}_{-0.04}$\\
	Eccentricity $e$                &                      & 0 (fixed)                              & - \\
	Argument of                     & ($^{\circ}$)         & 0 (fixed)                              & - \\
	periastron $\omega$ & & & \\
	\midrule
	\multicolumn{4}{l}{\textbf{Star}} \\
	Mass $M_\star$                    & (\mbox{M$_{\odot}$}) & $\mathcal{N_B}(0.88, 0.02, 0.0, 3.0)$  & $0.88 \pm 0.02$ \\
	Radius $R_\star$                  & (\mbox{R$_{\odot}$}) & $\mathcal{N_B}(0.87, 0.03, 0.0, 3.0)$  & $0.87^{+0.03}_{-0.02}$ \\
	\midrule	
	\multicolumn{4}{l}{\textbf{Photometry}} \\
	{\it TESS}$_{S1,2}$ mean        &                      & $\mathcal{N}(0.0, 1.0)$                & $0.00006 \pm 0.00001$ \\
	{\it TESS}$_{S28}$ mean         &                      & $\mathcal{N}(0.0, 1.0)$                & $0.00008 \pm 0.00002$ \\
	LCO$_1$ mean                    &                      & $\mathcal{N}(0.0, 1.0)$                & $0.00050 \pm 0.00007$ \\
	LCO$_2$ mean                    &                      & $\mathcal{N}(0.0, 1.0)$                & $0.00049 \pm 0.00007$ \\
	LCO$_3$ mean                    &                      & $\mathcal{N}(0.0, 1.0)$                & $0.0008 \pm 0.0001$ \\
	LCO$_4$ mean                    &                      & $\mathcal{N}(0.0, 1.0)$                & $0.00051 \pm 0.00006$ \\
	LCO$_5$ mean                    &                      & $\mathcal{N}(0.0, 1.0)$                & $0.00053 \pm 0.00009$ \\
	LCO$_6$ mean                    &                      & $\mathcal{N}(0.0, 1.0)$                & $0.0005 \pm 0.0001$ \\
	\midrule
	\multicolumn{4}{l}{\textbf{HARPS RVs}} \\
	$\log{(K)}$                     &       & $\mathcal{U}(0.0, 10.0)$               & $3.77 \pm 0.02$ \\
	Offset                          & (m\,s$^{-1}$)        & $\mathcal{U}(-6702, -6682)$          & $-6692.4 \pm 0.8$ \\
	$\log{(\rm{Jitter)}}$           &       & $\mathcal{N}(2.193774^{\dagger}, 5.0)$ & $-1.1^{+1.1}_{-1.8}$\\
	\bottomrule
	\end{tabular}
	\begin{tablenotes}
	\item \textbf{Distributions:}
	\item $\mathcal{N}(\mu, \sigma)$: a normal distribution with a mean $\mu$ and a standard deviation $\sigma$;
	\item $\mathcal{N_B}(\mu, \sigma, a, b)$: a bounded normal distribution with a mean $\mu$, a standard deviation $\sigma$, a lower bound $a$, and an upper bound $b$ (bounds optional);
	\item $\mathcal{U}(a, b)$: a uniform distribution with a lower bound $a$, and an upper bound $b$.
	\item \textbf{Prior values:}
	\item $^*$ equivalent to $0.5(\log{(D)}) + \log{(R_\star)} \pm 1$ where $D$ is the transit depth (ppm multiplied by $10^{-6}$) and $R_\star$ is the mean of the prior on the stellar radius (\mbox{R$_{\odot}$}), and $-1$ computes the lower bound while $+1$ computes the upper bound;
	\item $^{\dagger}$ equivalent to the log of the minimum error on the HARPS data (m\,s$^{-1}$).
	\end{tablenotes}
	\end{threeparttable}
\end{table*}

\section{Co-orbital bodies}

The full set of priors and posteriors from each of the four co-orbital scenario models described in Section\,\ref{coorbital} are presented in Table \ref{tab:coorbitals}. 

\begin{table*}
\small
\caption{Co-orbital hypothesis results.}
\label{tab:coorbitals}
\begin{tabular}{p{1.25cm}p{3cm}llll}
\toprule
\multirow{2}{*}{Parameter}         & \multirow{2}{*}{Priors}   & \multicolumn{4}{c}{Posteriors} \\
                                   &                           & 1p(c)                 & 1p                          & 1p(c)T             & 1pT \\
\midrule
$P$ (days)                         & $\mathcal{G}$(0.77703817, 10$^{-8}$)              & $0.77703792\pm9.9\times10^{-7}$  & $0.77703812\pm9.9\times10^{-7}$ & Fixed & Fixed \\
$t_{\rm 0}$ (BJD-2457000)          & $\mathcal{G}$(2062.444852864292, 4$\times10^{-7}$) & $2062.44496903\pm4\times10^{-7}$ & $2062.44496903 \pm 4\times 10^{-7}$ & Fixed  & Fixed \\
$K_{\rm b}$ (m\,s$^{-1}$)          & $\mathcal{U}$(0.0,100.0)  & $43.1^{+1.2}_{-1.2}$ & $43.2^{+1.2}_{-1.2}$        & $43.1^{+1.2}_{-1.2}$& $43.2^{+1.2}_{-1.2}$ \\
$e$                                & $\mathcal{U}$(0.0,1.0)    & Circular             & $0.025^{+0.021}_{-0.016}$   & -                   & - \\
$\omega$ (deg.)                    & $\mathcal{U}$(0.0,359.99) & Circular             & $215^{+38}_{-72}$           & -                   & - \\
$c$                                & $\mathcal{G}$(0.0,0.05)   & -                    & -                           & -                   & $0.028\pm0.025$ \\
$d$                                & $\mathcal{G}$(0.0,0.05)   & -                    & -                           & -                   & $0.014\pm0.026$ \\
$\delta_{\rm HARPS}$ (km\,s$^{-1}$) & $\mathcal{U}$(-15.0,-0.0) & $-6.69198^{+0.00091}_{-0.00088}$ & $-6.69194^{+0.00088}_{-0.00088}$ & $-6.69209^{+0.00091}_{-0.00089}$ & $-6.69184^{+0.00091}_{-0.00090}$ \\
$\sigma_{\rm HARPS}$ (m\,s$^{-1}$) & $\mathcal{U}$(0.0,5.0)    & $1.6^{+1.3}_{-1.0}$  & $1.33^{+1.3}_{-0.91}$       & $1.5^{+1.4}_{-1.0}$ & $1.4^{+1.4}_{-1.0}$ \\
\midrule
$\ln{\mathcal{Z}}$                 &                           & 55.5                 & 52.7                        & 48.9                & 48.4 \\
\bottomrule
\end{tabular}
\begin{tablenotes}
\item Posteriors are given for the following models, as described in Section\,\ref{coorbital}: \\
``1p(c)'', one planet on a circular orbit; \\
``1p'', one planet with the possibility of an eccentric orbit; \\
``1p(c)T'', a co-orbital scenario with a circular orbit; \\
``1pT'', a co-orbital scenario with the possibility of an eccentric orbit.
\item The period and time of conjunction posteriors coincide with the priors as they are much more constrained by the transit modelling. All posteriors for the systemic velocity agree within 1$\sigma$.
\end{tablenotes}
\end{table*}

\clearpage

\section{Author affiliations}\label{sec:affiliations}

$^{1}$Department of Physics, University of Warwick, Gibbet Hill Road, Coventry CV4 7AL, UK\\
$^{2}$Centre for Exoplanets and Habitability, University of Warwick, Gibbet Hill Road, Coventry CV4 7AL, UK\\
$^{3}$Institute for Computational Science, University of Zurich, Winterthurerstr. 90, 8057 Zurich, Switzerland\\
$^{4}$Instituto de Astrof\'isica e Ci\^encias do Espa\c{c}o, Universidade do Porto, CAUP, Rua das Estrelas, 4150-762 Porto, Portugal\\
$^{5}$Center for Astrophysics \textbar \ Harvard \& Smithsonian, 60 Garden Street, Cambridge, MA 02138, USA\\
$^{6}$Department of Space, Earth and Environment, Chalmers University of Technology, Onsala Space Observatory, 439 92 Onsala, Sweden\\
$^{7}$Astrophysics Group, Keele University, Staffordshire ST5 5BG, UK\\
$^{8}$Astrophysics Research Centre, School of Mathematics and Physics, Queen’s University Belfast, BT7 1NN Belfast, UK\\
$^{9}$Department of Astronomy, University of Michigan, Ann Arbor, MI 48109, USA\\
$^{10}$Centro de Astrobiolog\'ia (CAB, CSIC-INTA), Depto. de Astrof\'isica, ESAC campus, 28692, Villanueva de la Ca\~nada (Madrid), Spain\\
$^{11}$U.S. Naval Observatory, Washington, D.C. 20392, USA\\
$^{12}$Max Planck Institute for Astronomy, Heidelberg, Germany\\
$^{13}$Departamento de F\'isica e Astronomia, Faculdade de Ci\^encias, Universidade do Porto, Rua do Campo Alegre, 4169-007 Porto, Portugal\\
$^{14}$Department of Physics and Astronomy, Vanderbilt University, Nashville, TN 37235, USA\\
$^{15}$Perth Exoplanet Survey Telescope\\
$^{16}$Department of Physics and Kavli Institute for Astrophysics and Space Research, Massachusetts Institute of Technology, Cambridge, MA 02139, USA\\
$^{17}$Department of Earth, Atmospheric and Planetary Sciences, Massachusetts Institute of Technology, Cambridge, MA 02139, USA\\
$^{18}$Department of Aeronautics and Astronautics, MIT, 77 Massachusetts Avenue, Cambridge, MA 02139, USA\\
$^{19}$Department of Astrophysical Sciences, Princeton University, Princeton, NJ 08544, USA\\
$^{20}$NASA Ames Research Center, Moffett Field, CA 94035, USA\\
$^{21}$Department of Astronomy, MC 249-17, California Institute of Technology, Pasadena, CA 91125, USA\\
$^{22}$NASA Exoplanet Science Institute - Caltech/IPAC, Pasadena, CA USA\\
$^{23}$George Mason University, 4400 University Drive, Fairfax, VA, 22030 USA\\
$^{24}$NASA Goddard Space Flight Center, Exoplanets and Stellar Astrophysics Laboratory (Code 667), Greenbelt, MD 20771, USA\\
$^{25}$International Center for Advanced Studies (ICAS) and ICIFI (CONICET), ECyT-UNSAM, Campus Miguelete, 25 de Mayo y Francia, (1650) Buenos Aires, Argentina\\
$^{26}$ETH Zurich, Institute for Particle Physics and Astrophysics, Wolfgang-Pauli-Strasse 27, CH-8093 Zurich Switzerland\\
$^{27}$Department of Astronomy of the University of Geneva, Geneva Observatory, Chemin Pegasi 51, 1290 Versoix, Switzerland\\
$^{28}$Department of Astronomy and Tsinghua Centre for Astrophysics, Tsinghua University, Beijing 100084, China\\
$^{29}$European Southern Observatory, Karl-Schwarzschild-Stra{\ss}e 2, 85748 Garching bei M{\"u}nchen, Germany\\
$^{30}$Physikalisches Institut, University of Bern, Gesellsschaftstrasse 6, 3012 Bern, Switzerland\\
$^{31}$South African Astronomical Observatory, P.O. Box 9, Observatory, Cape Town 7935, South Africa\\
$^{32}$SETI Institute, Mountain View, CA 94043 USA\\

\bsp	
\label{lastpage}
\end{document}